\documentclass[aps,preprint]{revtex4}

\usepackage{amsfonts}
\usepackage{amssymb}
\usepackage{amsmath}
\usepackage{graphicx}
\usepackage{upgreek}
\usepackage{float}
\usepackage{dsfont}
\usepackage{multirow}
\usepackage[sort&compress]{natbib}

\def\dint{\displaystyle \int }

\def\dsum{\mathop{\displaystyle \sum }}

\begin{document}

\title{One-particle irreducible functional approach
-- a new route to diagrammatic extensions of DMFT}
\author{G. Rohringer$^{a}$,  A. Toschi$^{a}$,  H.
Hafermann$^{b}$, K. Held$^{a}$, V.I. Anisimov$^{c,d}$, A. A. Katanin$^{c,d}$}
\affiliation{
$^{a}$Institute for Solid State Physics, Vienna University of Technology, 1040 Vienna, Austria\\
$^{b}$Centre de Physique Th\'eorique, \'Ecole Polytechnique, CNRS, 91128 Palaiseau Cedex, France\\
$^{c}$Institute of Metal Physics, 620990, Ekaterinburg, Russia\\
$^{d}$Ural Federal University, 620002, Ekaterinburg, Russia }

\begin{abstract}
We present an approach which is based on the one-particle irreducible (1PI) generating functional formalism and
includes electronic correlations on all length-scales beyond the local correlations of dynamical mean field theory (DMFT).
This formalism allows us to unify aspects of the dynamical vertex approximation (D$\Gamma$A) and the dual fermion (DF)
scheme, yielding a consistent formulation of non-local correlations at the one- and two-particle level beyond DMFT
within the functional integral formalism.
In particular, the considered approach includes one-particle reducible contributions from the three- and more-particle vertices in the
dual fermion approach, as well as some diagrams not included in the ladder version of D$\Gamma$A. To demonstrate the applicability and physical
content of the 1PI approach, we compare the diagrammatics of 1PI, DF and D$\Gamma$A, as well as the
numerical results of these approaches for the
half-filled Hubbard model in two dimensions.
\end{abstract}

\maketitle

\section{Introduction}

Dynamical mean-field theory (DMFT) \cite{DMFT1,DMFT2}   represents  a big step forward for our understanding of strongly correlated electron systems. It fully includes local correlations, which often constitute the major contribution of electronic correlations. These are crucial for quasiparticle renormalization and the physics of the Mott-Hubbard metal-insulator transition (MIT)\cite{MH}. However, the arguably most fascinating phenomena of electronic correlations, such as unconventional superconductivity or (quantum) criticality,  originate from (or at least are strongly affected by) {\sl nonlocal} correlations.
Hence, several approaches have been developed using DMFT as a starting point and including nonlocal correlations beyond. The two main routes to
this end are cluster \cite{DCA,LK,clusterDMFT,clusterreview}
and diagrammatic extensions \cite{Schiller,DGA,DGAsimil,DF,LDFA} of DMFT.

The basic idea of cluster extensions is to go beyond the single-site DMFT by extending the notion of locality to a
cluster of sites. This way, correlations on length scales of the extension of the cluster, which hence are
short-ranged, are included. In practice, two different flavors are employed, which are essentially based on clusters in
real- or ${\bf k}$-space, and are coined cellular DMFT\cite{clusterDMFT} and dynamical cluster approximation
(DCA)\cite{DCA}, respectively. Complementarily, two distinct diagrammatic extensions of DMFT, based on the computation
of the local two-particle vertex\cite{localvertexinfo} of the Anderson impurity model (AIM) associated to DMFT, have
been proposed. Both diagrammatic schemes aim at the inclusion of short- {\em and}  long-range  nonlocal correlations, and share, to some extent, a similar philosophy\cite{noteJanis} with the
diagrammatic treatments of the Anderson localization built around the infinite dimensional limit\cite{Janis1, Janis2}.
The two diagrammatic extensions of DMFT, however, differ:  the dynamical vertex approximation  (D$\Gamma$A)\cite{DGA,DGA2,DGAsimil} is based on
the consideration of the fully {\sl two}-particle \emph{irreducible} local vertex, while the dual fermion (DF)\cite{DF,LDFA, DFparquet}
diagrams are built from the two-particle local vertex which is one- and two-particle \emph{reducible}.

The idea behind D$\Gamma$A is a systematic resummation of the most relevant Feynman diagrams beyond the DMFT ones:
While DMFT is based on the locality of the fully irreducible one-particle vertex (i.e., the self-energy), D$\Gamma$A
raises this locality concept to a higher level of the diagrammatics, requiring only the fully irreducible $n$-particle
vertex to be local. Fortunately, there is compelling numerical evidence that, even in two dimensions, the {\em fully
irreducible} $n\!=\!2$-particle vertex indeed is local\cite{Maier06}, so that this vertex can be considered as a building
block of the diagram technique in the D$\Gamma$A approach.
The proper D$\Gamma$A treatment would hence correspond to the solution of the parquet equations\cite{parquet}, with an
input given by the two-particle irreducible local vertex function.  While the numerical solution of the parquet
equations has been recently achieved\cite{parquetsolv} for single-band two-dimensional models, the computational effort
is still considerable. Hence, most of the D$\Gamma$A results obtained hitherto\cite{DGA2,DGA3} employed the ladder approximation, where,
taking into account the most important channel(s) only, the assumption of locality is made for the two-particle vertices, which are irreducible in these channels.

The DF approach, instead, is a systematic functional-integral expansion around DMFT.
By introducing an impurity problem at each lattice site, the lattice action is recast in terms of decoupled impurities and a momentum-dependent remainder, which involves the hybridization function and the bare dispersion. Through a Hubbard-Stratonovich transformation acting on this term, so called dual fermions are introduced. These couple only \emph{locally} to the original fermionic degrees of freedom.
Hence, the latter can be integrated out, which produces {\sl all} local connected two- and more-particle diagrams (vertices) of the impurity problem through which the dual fermions are coupled. This is in contrast with the D$\Gamma$A which uses only the irreducible part of these vertices\cite{Katanin12}.
Analogously to D$\Gamma$A, the inclusion and an exact
treatment of all $n$-particle interaction terms among the electrons would correspond to the exact solution of the
problem, but in practice three- and more-particle vertices are neglected and only the lowest
order interaction terms (i.e., the two-particle local vertex) for the dual electrons are considered. Different
diagrammatic approximations within the DF approach, such as second-order perturbation theory\cite{DF}, ladder series\cite{LDFA},
and quite recently even parquet\cite{DFparquet} have been considered.
We note here, that the DF parquet calculations, when performed without self-consistency, would be formally similar to
the parquet corrections to the local physics of the Anderson localization problem considered in Ref. \cite{Janis1},
though in the former case the local connected two-particle vertex is obviously much more complex than the one of CPA.

A thorough comparison between the diagrammatics of DF and D$\Gamma$A has not
been done so far, although, from the above discussion one may surmise a sort of underlying similarity between the two diagrammatic approaches and
their schemes of implementation. To perform such a comparison,
we present in this paper a general approach for a systematic
inclusion of nonlocal corrections beyond DMFT. This new scheme is also based on a functional integral, similarly as in
the DF approach, but it is formulated in terms of the one particle irreducible (1PI) vertices instead of the reducible
vertices of the DF approach. In order to illustrate the content of the 1PI approach we  compare it
diagrammatically and numerically with DF and D$\Gamma$A, also illustrating
 the diagrammatic
relations between the latter two approaches. Quite remarkably, our results demonstrate that the 1PI approach
combines
synergetically important features of the DF and D$\Gamma$A schemes.

In Section II we  discuss the general structure of the nonlocal corrections to DMFT, considering contributions to the
self-energy which are second-order with respect to the bare on-site Coulomb repulsion and the DMFT local interaction,
respectively. In Section III we develop a new formalism based on the transformation to the one-particle irreducible
functional in the DMFT-part of the action. In Section IV we derive  nonlocal contributions to the local (DMFT)
self-energy based on ladder diagrams and discuss these in terms of a comparison with the DF and D$\Gamma$A approaches.
In Section V, we discuss results for the two-dimensional Hubbard model obtained with our new method, and, finally,
Section VI is devoted to conclusions and an outlook.

\section{\textbf{Second-order perturbation theory}}

Let us briefly discuss the structure of the corrections to the dynamical mean-field theory by means of
the perturbation theory.
Specifically, we consider  the Hubbard model
with hopping $t_{ij}$ and Coulomb interaction $U$:
\begin{equation}
H=\sum_{ij\sigma }t_{ij}\hat{c}_{i\sigma }^{\dagger }\hat{c}_{j\sigma
}+U\sum_{i}\hat{n}_{i\uparrow }\hat{n}_{i\downarrow }.  \label{H}
\end{equation}%
Here, the operator $\hat{c}_{i\sigma }^{\dagger }$  ($\hat{c}_{i\sigma}$) creates (annihilates)
an electron with spin $\sigma$ at lattice-site $i$, $\hat{n}_{i \sigma}=\hat{c}_{i\sigma }^{\dagger }\hat{c}_{i\sigma}$.
For the sake of simplicity,
this paper deals with the one-band Hubbard model only, but a generalization
of the 1PI approach to more complex multi-orbital models is, as a matter of course, possible.

The dynamical mean-field theory approximates the corresponding
full action by an effective local action \cite{DMFT2}
\begin{equation}
\mathcal{S}_{\text{DMFT}}[c^+,c]=-\sum_{i}\frac{1}{\beta^2}\dint\limits_{0}^{\beta } d\tau \dint\limits_{0}^{\beta } d\tau' \dsum\limits_{\sigma }c_{i\sigma }^{+}(\tau ) \zeta^{-1}(\tau-\tau')
c_{i\sigma}(\tau')+ \dint\limits_{0}^{\beta } d\tau\; Un_{i\uparrow }(\tau )n_{i\downarrow }(\tau ).
\label{LDMFT}
\end{equation}
where $c_{i\sigma}^{+}(\tau )$
and $c_{i\sigma }(\tau )$ are Grassmann variables
corresponding to the Fermion operators  $\hat{c}_{i\sigma }^{\dagger }$ and $\hat{c}_{i\sigma}$
at imaginary time $\tau$, $\beta=1/T$ is the inverse temperature.
The "Weiss field", i.e., the non-interacting impurity Green's function $\zeta (\tau-\tau')$, has to be determined
self-consistently in DMFT from the following condition on its Fourier transform $\zeta_{\nu}$
\begin{equation}
\dsum\limits_{\mathbf{k}}G_k=\left(\zeta ^{-1}_{\nu}-\Sigma_{\text{loc},\nu}\right)^{-1}=G_{\text{loc},\nu}^{-1}
\label{sc}
\end{equation}
where
\begin{equation}
G_k=\left( i\nu+\mu-\varepsilon_{\mathbf{k}}-\Sigma_{\text{loc},\nu}\right)^{-1},  \label{Glattice}
\end{equation}
$\varepsilon_{\mathbf{k}}\,$\ is the Fourier transform of $t_{ij}$, $\mu$ is the chemical potential, and $\Sigma_{\text{%
loc},\nu}$ is the self-energy of the impurity problem [see Eq.\ (\ref{LDMFT})]
at the fermionic Matsubara frequency $i\nu$ [i.e., $\nu=\frac{\pi}{\beta}(2n+1),n\in\mathds{Z}$]. Note that we specify all imaginary frequency arguments as subscripts (or, for the vertex functions below, as superscripts) and that we adopt a four-vector notation for the frequency and momentum arguments, i.e., $k=(\nu,\mathbf{k})$ for a fermionic and $q=(\omega,\mathbf{q})$ for a bosonic Matsubara frequency [i.e., $\omega=\frac{\pi}{\beta}(2m),m\in\mathds{Z}$]. The subscript ``loc'' is attached to all quantities (Green's functions and vertices) of the local AIM despite the Weiss fields $\zeta_{\nu}$. In practice,
the local problem in Eq.\ (\ref{LDMFT}) is solved numerically by exact
diagonalization or  quantum Monte-Carlo simulations \cite{DMFT2}, yielding a self-energy $\Sigma_{\text{loc},\nu}$ and Green's function $G_{\text{loc},\nu}$ until self-consistency regarding Eq.\ (\ref{sc}) is obtained. Since such numerical calculations can be better performed in Matsubara frequencies, we stick to
this formalism in the following, but a transformation to real frequencies is possible.

DMFT takes into account local dynamical correlations but it neglects inter-site correlations, which is reflected in a wave-vector $\mathbf{k}$ independent self-energy $\Sigma_k\equiv\Sigma_{\text{loc},\nu}$.
Perturbation theories such as self-consistent T-matrix,
fluctuation exchange (FLEX) and parquet approximation   \cite{parquet2} can result in a $\mathbf{k}$-dependent $\Sigma $, but the most important local correlations are not reliably reproduced when the system is not in the weak coupling regime, i.e., if  the
 Coulomb interaction parameter $U$ is comparable to or larger than the band width.

To illustrate the structure of nonlocal corrections beyond DMFT, we first analyze the
perturbation theory. Since we want to find corrections to the already calculated local (DMFT) self-energy we use the DMFT Green's function, given in Eq. (\ref{Glattice}), as ``bare'' propagator for the construction of  self-energy diagrams. Let us now separate purely local and nonlocal contributions to $\Sigma_k$ by introducing the function
\begin{equation}
\widetilde{G}_k\equiv G_k-G_{\text{loc},\nu},  \label{F5}
\end{equation}
which vanishes after averaging in $\mathbf{k}$ space by construction:
\begin{align}
\sum_{\mathbf{k}}\widetilde{G}_k=0.\label{F6}
\end{align}
In the second order in $U$ we obtain for the non-local self-energy:
\begin{align}
\Sigma ^{(2)}_k&=\frac{U^2}{\beta}\sum_{q}G_{k-q}V^{(2)}_q=
\Sigma _{\text{loc},\nu}^{(2)}+\widetilde{\Sigma }^{(2)}_k  \label{F12} \\
\Sigma _{\text{loc},\nu}^{(2)}&=\frac{U^2}{\beta}\sum_{\omega}G_{\text{loc},\nu-\omega}V_{\text{loc},\omega}^{(2)}
\nonumber \\
\widetilde{\Sigma }^{(2)}_k&=\frac{U^2}{\beta}\sum_{q}\widetilde{G}_{k-q}\widetilde{V}^{(2)}_q,  \nonumber
\end{align}%
where
\begin{align}
V^{(2)}_q&=-\frac{1}{\beta}\sum_{k'}G_{k'+q}G_{k'}  =V_{\text{loc},\omega}^{(2)}+\widetilde{V}^{(2)}_q , \label{F4} \\
V_{\text{loc},\omega}^{(2)}&=\sum_{\nu'}\chi_{\text{loc}}^{0,\nu^{\prime}\omega},\quad
\widetilde{V}^{(2)}_q=\sum_{\nu'}\widetilde{\chi}_q^{\nu^{\prime}}
 \notag,
\end{align}%
and $\chi^{0,\nu\omega}_{\text{loc}}$ and $\widetilde{\chi}_q^{\nu}$ are defined as:
\begin{align}
\chi_{\text{loc}}^{0,\nu\omega}&=-\frac{1}{\beta}G_{\text{loc},\nu}G_{\text{loc},\nu+\omega}, \\
\widetilde{\chi}_q^{\nu} &=-\frac{1}{\beta}\sum_{\mathbf{k}}\widetilde{G}_k\widetilde{G}_{k+q}.  \notag
\end{align}
The ``mixed'' local-nonlocal terms in Eq.\ (\ref{F12}) vanish due to the identity in Eq.\ (\ref{F6}). For the same reason $\widetilde{V}
^{(2)}_q$ vanishes after averaging in $\mathbf{k}$ space:
\begin{equation}
\sum_{\mathbf{q}}\widetilde{V}^{(2)}_q=-\frac{1}{\beta}\sum_{
k'}\left\{ \sum_{\mathbf{q}}\widetilde{G}_{k'+q}\right\} \widetilde{G}_{k'}=0.  
\end{equation}
The local part $\Sigma _{\text{loc},\nu}$ in Eq.\ (\ref{F12}) can be replaced by its DMFT value, so that only nonlocal terms are calculated by perturbation theory.

\begin{figure}[t]
 \centering
 \includegraphics[width=0.5\textwidth]{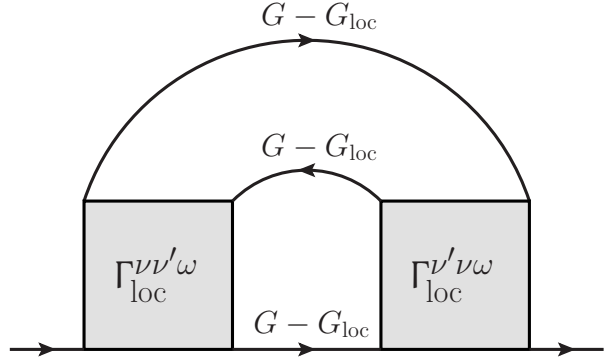}
 \caption{2$^{\text{nd}}$-order diagram for the DF approach in terms of real electrons.}
 \label{fig:DF2ndord}
\end{figure}

In  higher orders of the perturbation theory, there is no such clear separation of terms: mixed local-nonlocal terms appear in
 $\Sigma_k$. Considering, however, the leading
nonlocal correction to the local self-energy, these terms can be reduced to those
containing the \textit{local} vertex instead of $U$ in the second-order result, Eq.\ (\ref{F12}), and the \textit{nonlocal} part of the
Green's functions. In particular, using the dual fermion approach\cite{DF}
the corresponding correction can be expressed as
\begin{equation}
\Sigma ^{(2)}_{\text{d},k}=\frac{1}{2\beta}\sum\limits_{\nu',q}\sum\limits_{m=c,s}A_{m}\Gamma _{\text{loc},m}^{\nu \nu ^{\prime } \omega }\widetilde{\chi }^{\nu ^{\prime }}_q \Gamma _{\text{loc},m}^{\nu ^{\prime
 } \nu \omega } G_{k+q}\label{St},
\end{equation}
where  $A_{s}=3/2;A_{c}=1/2$,
$\Gamma _{\text{loc},s(c)}^{\nu \nu ^{\prime }\omega }=-\Gamma _{\text{loc},\uparrow \uparrow}^{\nu \nu ^{\prime }\omega}\pm \Gamma _{\text{loc},\uparrow \downarrow}^{\nu \nu ^{\prime }\omega}$ is the
local two-particle vertex, which is related to the local susceptibility
\begin{eqnarray}
\chi _{\text{loc},\sigma\sigma'}^{\nu \nu ^{\prime }\omega }&=&\frac{1}{\beta^2}\int_{0}^{\beta}d\tau _{1}\,d\tau _{2}\,d\tau _{3}\;\mbox{e}%
^{-i\tau _{1}\nu }\,\mbox{e}^{i\tau _{2}(\nu +\omega )}\,\mbox{e}^{-i\tau
_{3}(\nu ^{\prime }+\omega )}  \label{Eq:chiloc} \\
&\times &\left[ \langle T_{\tau }\;\hat{c}_{i\sigma }^{\dagger }(\tau
_{1})\hat{c}_{i\sigma }(\tau _{2})\hat{c}_{i\sigma ^{\prime }}^{\dagger }(\tau
_{3})\hat{c}_{i\sigma ^{\prime }}(0)\rangle \right.  \notag \\
&-&\left. \langle T_{\tau }\;\hat{c}_{i\sigma }^{\dagger }(\tau _{1})\hat{c}_{i\sigma
}(\tau _{2})\rangle \langle T_{\tau }\hat{c}_{i\sigma ^{\prime }}^{\dagger }(\tau
_{3})\hat{c}_{i\sigma ^{\prime }}(0)\rangle \right]   \notag
\end{eqnarray}
by
\begin{equation}
\Gamma _{\text{loc},\sigma \sigma^{\prime}}^{\nu \nu ^{\prime }\omega }=-
\frac{\chi _{\text{loc},\sigma \sigma^{\prime}}^{\nu \nu ^{\prime }\omega }-\chi^{0,\nu \omega}_{\text{loc}}\delta _{\nu \nu ^{\prime }}\delta _{\sigma \sigma ^{\prime }}}{\chi^{0,\nu \omega}_{\text{loc}}\chi
^{0,\nu ^{\prime }\omega}_{\text{loc}}} . \label{Gamma_loc} \;
\end{equation}%
The susceptibilities
$\chi _{\text{loc},\sigma\sigma^{\prime}}^{\nu \nu ^{\prime }\omega }$
can be obtained from the exact diagonalization or quantum Monte Carlo
solution of the single-impurity problem.
The result (\ref{St}) is illustrated diagrammatically in Fig. \ref{fig:DF2ndord}.

\begin{figure}
 \includegraphics[width=0.8\textwidth]{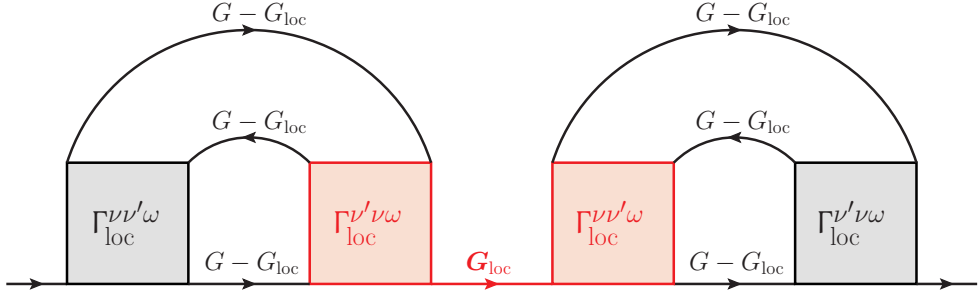}
 \caption{A reducible diagram for the real self-energy $\Sigma_k$ stemming from the expansion of the denominator in Eq.\ (\ref{Eq:SDF}) and the diagram of Fig.\ \ref{fig:DF2ndord} as numerator. In the DF approach, the same contribution, albeit with opposite sign, is generated from a diagram involving the three-particle vertex, which contains the part marked in red, hence canceling this reducible contribution.}
 \label{fig:DFreducible}
\end{figure}

In the DF  approach the self-energy $\Sigma_{\rm d,k}$ is however an auxiliary construct. It is related to the real self-energy
$\Sigma_k$ of the system via
\begin{equation}
\label{Eq:SDF}
\Sigma_k=\Sigma_{\text{DF},k}=\frac{\Sigma_{\text{d},k}}{1+G_{\text{loc},\nu}\Sigma_{\text{d},k
}} + \Sigma_{\text{loc},\nu} \; .
\end{equation}
In order to understand this equation diagrammatically, one can expand the denominator in a geometric series. This
procedure generates, together with $\Sigma_{\rm d}$ from Fig.\ \ref{fig:DF2ndord}, the additional diagram depicted in
Fig.\ \ref{fig:DFreducible} and chain expansions thereof. Evidently, the lattice self energy
should not contain one-particle reducible Feynman diagrams. Indeed,
as discussed in Ref. \cite{Katanin12}, including three- and more-particle vertices in the DF calculations actually removes these spurious
contributions to the self-energy. In our example, the diagram shown in Fig.\ \ref{fig:DFreducible}
is  canceled by a corresponding contribution from the one-particle reducible three-particle  vertex
(shown in red in Fig.\ \ref{fig:DFreducible}).

The above discussed difficulty is obviously not a property of the DF approach per se but its truncation at the two-particle vertex level
while keeping, at the same time, the full denominator of  Eq.\ (\ref{Eq:SDF}).
This is analogous to the linked cluster theorem, as a consequence of which all vacuum to vacuum diagrams cancel in the ratio of path integrals which appears in the calculation of correlation functions.
Of course, this is no longer true if we perform an expansion of  the path integral in the numerator only up to a certain order.
Similarly, if we truncate the DF approach at the two-particle vertex level, reducible diagrams
stemming from local one-particle reducible three- and more-particle vertex functions are not present and, hence, the
denominator in Eq. (\ref{Eq:SDF}) introduces such terms in the diagrammatic expansion rather than canceling them. To
avoid these complications as well as to account systematically for the contribution of one-particle reducible diagrams
to three- and more-particle vertices, we consider below the one-particle irreducible formulation of the generating functional approach.

\section{The one-particle irreducible (1PI) approach}

For a general formulation of the nonlocal corrections to DMFT we separate the
nonlocal degrees of freedom in the generating functional formalism. To this
end, we consider the generating functional%
\begin{equation}
Z[\eta^+,\eta]\! =\!\! \dint\!  D[c^+,c]\exp \left\{\!  -\!\!   \int_0^{\beta}\!\!  d\tau\!\left[
\sum_{i,\sigma}\!  {\left( c^{+}_{i\sigma}(\!\tau\!) \frac{\partial c_{i\sigma}(\!\tau\!) }{\partial \tau }\! -\! \eta ^{+}_{i\sigma}(\!\tau\!)c_{i\sigma}(\!\tau\!)\! -\! c^{+}_{i\sigma}(\!\tau\!)
\eta_{i\sigma}(\!\tau\!) \right)}\!  +\! H[c^+,c]\right]\!  \right\},  \label{gen}
\end{equation}%
where $\eta_{i \sigma}^+(\tau) ,\eta_{i \sigma}(\tau)$ are the fermionic source fields.
The contribution of the local and nonlocal degrees of freedom can be split by
performing a Fourier transform in the exponent and introducing the
auxiliary fields $\widetilde{c}^+$ and $\widetilde{c}$ \cite{noteequiv}:%
\begin{eqnarray}
Z[\eta^+,\eta] &=&\dint D[\widetilde{c}^+,\widetilde{c}]\exp \left\{%
\beta \sum_{k,\sigma}\widetilde{c}^{+}_{k\sigma}\left(\zeta_{\nu}^{-1}-G_{0k}^{-1}\right)^{-1}%
\widetilde{c}_{k\sigma}\right\}\label{Split}\\
&&\hspace{-0.6cm}\times \dint D[c^+,c]\exp \left\{ -\mathcal{S}_{\text{DMFT}%
}[c^+,c]+\sum_{k,\sigma}\left[(\eta_{k\sigma} ^{+}+
\widetilde{c}_{k\sigma}^{+})c_{k\sigma}+c_{k\sigma}^{+}(\eta_{k\sigma} +%
\widetilde{c}_{k\sigma})\right]\right\}, \notag
\end{eqnarray}%
where $G_{0k}^{-1}=i\nu-\varepsilon_{\mathbf{k}}$ is the non-interacting lattice Green's function. Let us recall that the correlation (or Green's) functions can be obtained by functional derivatives of $\log Z[\eta^+,\eta]$ with respect to $\eta^+$ and $\eta$ at $\eta^+=\eta=0$, which allows us to neglect any normalization factor which would appear in front of the integral on the right hand side of Eq. (\ref{Split}).

Whereas up to this point the formalism is essentially the same as in the
derivation of the DF approach \cite{DF,dgadfdifference},
we now apply a Legendre transform
in order to pass to the 1PI functional in the DMFT part of the action
\begin{eqnarray}
\exp (-W_{\text{DMFT}}[\widetilde{\eta}^+,\widetilde{\eta}]) &=&\dint D[c^+,c]\exp \left\{ -%
\mathcal{S}_{\text{DMFT}}[c^+,c]+\sum_{k,\sigma}\left(\widetilde{\eta }_{k\sigma}^{+}c_{k\sigma}+c^{+}_{k\sigma}\widetilde{%
\eta }_{k\sigma}\right)\right\}  \notag \\
&=&\exp \left\{-\Gamma _{\text{DMFT}}[\phi^+,\phi]+\sum_{k,\sigma}
\left( \widetilde{\eta}_{k\sigma} ^{+}\phi_{k\sigma} +\phi^{+}_{k\sigma}\widetilde{\eta}_{k\sigma}\right)\right\}  \label{gloc}
\end{eqnarray}%
where%
\begin{align}
&\phi_{k\sigma} =-\frac{\delta W_{\text{DMFT}}[\widetilde{\eta}^+,\widetilde{\eta }]%
}{\delta \widetilde{\eta }^{+}_{k\sigma}},&&\widetilde{\eta}_{k\sigma} =\frac{\delta \Gamma _{\text{DMFT}%
}[\phi^+,\phi]}{\delta \phi ^{+}_{k\sigma}}\ ,  \label{legendre}
\end{align}
$\widetilde{\eta }_{k\sigma}=\eta_{k\sigma} +\widetilde{c}_{k\sigma}$, and similarly [but with reversed sign in Eq. (\ref{legendre})] for the
conjugated fields. Therefore, Eq.\ (\ref{Split}) becomes%
\begin{eqnarray}
Z[\eta^+,\eta] &=&\dint D[\widetilde{c}^+,\widetilde{c}]\exp \left\{%
\beta\sum_{k,\sigma}\widetilde{c}^{+}_{k\sigma}\left(\zeta_{\nu}^{-1}-G_{0k}^{-1}\right)^{-1}%
\widetilde{c}_{k\sigma}\right. \notag\\
&&\left. +\sum_{k,\sigma}\left[(\eta_{k\sigma} ^{+}+\widetilde{c}_{k\sigma}%
^{+})\phi_{k\sigma} +\phi_{k,\sigma} ^{+}(\eta_{k\sigma} +\widetilde{c}_{k\sigma})\right]
-\Gamma _{\text{DMFT}}[\phi^+,\phi]\right\}.
\end{eqnarray}%
The fields $\phi^+$ and $\phi$ in this representation are functionals of
the fields $\widetilde{\eta}^+$ and $\widetilde{\eta }$ defined via the relations in Eq.\ (%
\ref{legendre}). For the following consideration it is convenient to change
the variables of integration from $\widetilde{c}^+,\widetilde{c}$ to $\phi^+,\phi$. This yields%
\begin{equation}
\begin{split}
Z[\eta^+,\eta]=&\!\!\displaystyle\int\! D[\phi^+,\phi]\\ &\times\exp  \left\{\!
\beta\sum_{k,\sigma}\left( \frac{\delta \Gamma _{\text{DMFT}}[\phi^+,\phi]}{%
\delta \phi _{k\sigma }}\!+\!\eta _{k\sigma }^{+}\!\right) [\zeta_{\nu}^{-1}\!-\!G_{0k}^{-1}]^{-1}\left(\! -\frac{\delta \Gamma _{\text{DMFT}}[\phi^+
,\phi]}{\delta \phi _{k\sigma }^{+}}\!+\!\eta _{k\sigma }\!\right) \right.
\\
&\hspace{1.2cm} \left. -\sum_{k,\sigma}\left(\frac{\delta \Gamma _{\text{DMFT}}[\phi^+
,\phi]}{\delta \phi _{k\sigma }}\phi_{k\sigma }-\phi
_{k\sigma }^{+}\frac{\delta \Gamma
_{\text{DMFT}}[\phi^{+},\phi]}{\delta \phi _{k\sigma }^{+}}\right) -\Gamma _{\text{DMFT}}[\phi^+,\phi]\right\}
J[\phi^{+},\phi], \label{Eq1PI}
\end{split}
\end{equation}%
where $J^{-1}[\phi^+,\phi]=\det \delta ^{2}\Gamma_{\text{DMFT}}/(\delta \phi ^{+}\delta \phi)$ is the determinant of the Jacobian of the corresponding
transformation, see Appendix A for more details.

We proceed now by expanding the functional $\Gamma_{\text{DMFT}}[\phi^+,\phi]$ into a series with respect to the
source fields $\phi^+$ and $\phi$. In the DF approach such
an expansion is performed for the functional $W_{\text{DMFT}}[\eta^+,\eta]$ which generates connected but in general one-particle
reducible vertex functions as the coefficients of this expansion. Expanding $\Gamma_{\text{DMFT}}$ instead,
one obtains the (local) one-particle irreducible vertex functions amputated by the outer legs.
Neglecting the constant zeroth order contribution, the resulting expansion up to fourth order, i.e., up to the level of the two-particle vertex function, reads
\begin{equation}
\Gamma _{\text{DMFT}}[\phi^+,\phi]=-\frac{1}{\beta}\dsum\limits_{k,\sigma}G_{\text{loc},\nu}^{-1}\phi _{k\sigma }^{+}\phi _{k\sigma }+\frac{1}{2 \beta^3}
\dsum\limits_{kk'q}\sum_{\sigma\sigma'}\widetilde{\Gamma}_{\text{loc},\sigma \sigma'}^{\nu\nu'\omega}\;(\phi _{k\sigma }^{+}\phi
_{k+q,\sigma})(\phi_{k'+q,\sigma'}^{+}\phi_{k'\sigma'}),  \label{GammaDMFT}
\end{equation}
where $\widetilde{\Gamma}^{\nu\nu'\omega}_{\text{loc},\sigma\sigma'}=\left(1-\frac{1}{2}\delta_{\sigma\sigma'}\right)\Gamma^{\nu\nu'\omega}_{\text{loc},\sigma\sigma'}$.

In the next step, we use the (approximate) DMFT functional $\Gamma_{\text{DMFT}}$ from Eq. (\ref{GammaDMFT})
for evaluating Eq. (\ref{Eq1PI}), i.e., we have to calculate the derivatives of
the functional $\Gamma_{\text{DMFT}}$ with respect to the fields $\phi^+$ and $\phi$.
While the formal derivation is given in Appendix A, let us here discuss the
most important features of the calculation. The exponent in Eq. (\ref{Eq1PI}) contains
a term proportional to $(\delta_{\phi}\Gamma_{\text{DMFT}})(\delta_{\phi^+}\Gamma_{\text{DMFT}})$
(where $\delta_{\phi}$ denotes the functional derivative w.r.t.\ the field $\phi$). Inserting now
the two-particle part of $\Gamma_{\text{DMFT}}$ into this expression clearly leads
to a term which is proportional to $(\Gamma_{\text{loc}})^2(\phi^+\phi)^3$.
Such a contribution stems from the local reducible three(and more)-particle  vertices, and is hence absent in the DF approach if we neglect these vertices.
At the same time, such contributions stemming from reducible (local) diagrams are fully taken into account in the 1PI approach when expanding $\Gamma_{\text{DMFT}}$ up to the two-particle level.
The above mentioned three-particle  contribution can be decoupled by another Hubbard-Stratonovich transformation  introducing new fields $\psi^+$ and $\psi$. The corresponding calculations are carried out in Appendix A and yield:
\begin{eqnarray}
 Z[\eta^+,\eta]&=&\int D[\phi^+,\phi]D[\psi^+,\psi]\;\exp\left\{\sum_{k,\sigma}\eta^+_{k\sigma}\left(\psi_{k\sigma}+\phi_{k\sigma}\right)+\left(\psi_{k\sigma}^++\phi^+_{k\sigma}\right)\eta_{k\sigma}\right. \nonumber\\
 \hspace{0.7cm}&+&\left. \frac{1}{\beta}\sum_{k,\sigma}G^{-1}_k\left(\phi_{k\sigma}^+\phi_{k\sigma}%
 +\psi_{k\sigma}^+\phi_{k\sigma}%
 +\phi_{k\sigma}^+\psi_{k\sigma}\right)+\left(G^{-1}_k-G^{-1}_{\text{loc},\nu}\right)\psi_{k\sigma}^+\psi_{k\sigma}%
 \right.\nonumber\\
 \hspace{1.9cm}&-&\frac{1}{\beta^3}\sum_{kk'q}\sum_{\sigma\sigma'}\widetilde{\Gamma}_{\text{loc},\sigma\sigma'}^{\nu\nu'\omega}\left[\left(\psi^+_{k\sigma}\phi_{k+q,\sigma}\right)\left(\phi^+_{k'+q,\sigma'}\phi_{k'\sigma'}\right)\right.\nonumber\\
 &+&\left.\left(\phi^+_{k\sigma}\phi_{k+q,\sigma}\right)\left(\phi^+_{k'+q,\sigma'}\psi_{k'\sigma'}\right)+\frac{1}{2}\left(\phi^+_{k\sigma}\phi_{k+q,\sigma}\right)\left(\phi^+_{k'+q,\sigma'}\phi_{k'\sigma'}\right)\bigr]\right\}J[\phi^+,\phi],
 \label{Zf}
\end{eqnarray}
where $G_k$ is defined by Eq. (\ref{Glattice}) and accounts for the local self-energy. Eq. (\ref{Zf}) expresses the
partition function through the one-particle irreducible local vertex $\Gamma_{\text{loc},\sigma\sigma'}^{\nu\nu'\omega}$
and the local self-energy, and represents one of the central results of the present paper.
The first line of Eq. (\ref{Zf}) includes the source fields, the second line contains the quadratic (``bare'') terms in fermionic fields,
and the last two lines correspond to the interaction between fermionic degrees of freedom.
The nonlocal  Green's functions of the lattice
 model  is defined as $\mathbb{G}_{k\sigma }= - \frac{1}{\beta}\langle \langle c_{k\sigma }| c_{k\sigma }^{+} \rangle \rangle$.
It can be calculated through derivatives of Eq.\ (\ref{Zf}) w.r.t. to the source fields $\eta^+$, $\eta$:
\begin{eqnarray}
\mathbb{G}_{k\sigma }&=&
 \frac{1}{\beta}\frac{\delta ^{2} \ln Z}{\delta \eta
_{k\sigma }^{+}\delta \eta _{k\sigma }} \notag \\
&=&- \frac{1}{\beta}\left[ \langle \langle \phi _{k\sigma }|\phi _{k\sigma
}^{+}\rangle \rangle+\langle \langle \phi _{k\sigma }|\psi _{k\sigma }^{+}\rangle \rangle
+\langle \langle \psi _{k\sigma }|\phi _{k\sigma
}^{+}\rangle \rangle+\langle \langle \psi _{k\sigma }|\psi _{k\sigma }^{+}\rangle \rangle\; \right] \label{equ:gdef}.
\end{eqnarray}
\begin{figure}[t]
 \centering
 \includegraphics[width=0.75\textwidth]{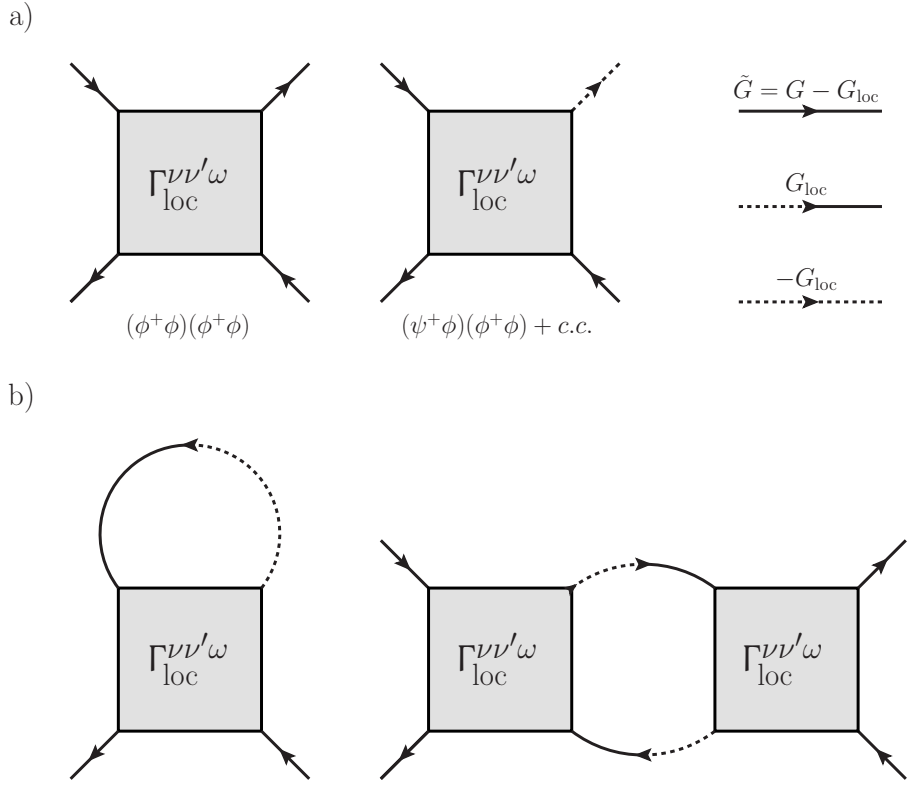}
 \caption{(a) Elements of the diagram technique in the 1PI approach. (b) Diagrams which are generated by the elements in (a) but excluded (canceled) by the corresponding counterterms in the (inverse) determinant $J[\phi^+,\phi]$ of the Jacobian (see Appendix A).}
 \label{fig:diagramelements1pi}
\end{figure}
That is, $\mathbb{G}_{k\sigma}$ can be written as the sum of four distinct propagators which can be combined in a more compact form
$-\frac{1}{\beta}\sum_{a,b=1,2} \langle \langle \Phi_{k\sigma}^a|\Phi _{k\sigma}^{+,b }\rangle \rangle$, where we have introduced a spinor
\begin{equation}
\Phi _{k\sigma }=\left(
\begin{tabular}{l}
$\phi _{k\sigma }$ \\
$\psi _{k\sigma }$
\end{tabular}
\right).
\end{equation}
In order to treat the interaction in Eq. (\ref{Zf}), we consider first the ''bare'' part in the action, which is quadratic in Grassmann variables. The corresponding 1PI ``bare'' propagators  are obtained by setting $\Gamma _{\text{loc}}\!=\!0$ in Eq. (\ref{Zf}) and yield
\begin{equation}
\mathcal{G}_{k}=-\frac{1}{\beta}\langle \langle \Phi _{k}|\Phi _{k}^{+ }\rangle \rangle_0
=\left(
\begin{array}{cc}
 G_{k}^{-1}& G_{k}^{-1} \\
G_{k}^{-1}& G_{k}^{-1}-G_{\text{loc},\nu}^{-1}
\end{array}
\right)^{-1}
=\left(
\begin{array}{cc}
G_{k}-G_{\text{loc},\nu} & G_{\text{loc},\nu} \\
G_{\text{loc},\nu} & -G_{\text{loc},\nu}
\end{array}
\right).  \label{G_nonint}
\end{equation}
Let us again stress that the ``bare'' functions $G_k$ and $G_{\text{loc},\nu}$
include {\sl all local} self-interaction effects via the local self-energy $\Sigma _{\text{loc},\nu}$, which is already considered in the DMFT part of the action [see Eqs. (\ref{Split}) and (\ref{gloc})].
The propagator $-\frac{1}{\beta}\langle \langle \phi _{k}|\phi _{k}^{+
}\rangle \rangle_{0}\equiv \widetilde{G}_{k}=G_k-G_{\text{loc},\nu}$ [as defined in Eq. (\ref{F5})] corresponds to the remaining nonlocal
fluctuations and obeys $\sum\nolimits_{\mathbf{k}}\widetilde{G}_k=0$. The propagators $-\frac{1}{\beta}\langle \langle \phi _{k}|\psi _{k}^{+ }\rangle \rangle_{0}=-\frac{1}{\beta}\langle \langle
\psi _{k}|\phi _{k}^{+ }\rangle \rangle_{0}=\frac{1}{\beta}\langle \langle \psi _{k}|\psi _{k}^{+
}\rangle \rangle_{0}=G_{\text{loc},\nu}$
describe the (``bare``) local quasiparticles, coupled to the nonlocal degrees of freedom
via the interaction in the third line in Eq. (\ref{Zf}).
The corresponding elements of the diagram technique are shown in Fig. \ref{fig:diagramelements1pi}a.
The interaction term consists of two parts which are depicted diagrammatically. The first diagram corresponds to the contribution $\Gamma_{\text{loc}}(\phi^+\phi)(\phi^+\phi)$ in the fourth line of Eq. (\ref{Zf}). This vertex can be either coupled to both local ($\langle\langle\phi|\psi^+\rangle\rangle$ or $\langle\langle\psi|\phi^+\rangle\rangle$) {\sl and} nonlocal propagators ($\langle\langle\phi|\phi^+\rangle\rangle$) or to nonlocal propagators {\sl only}.
In contrast, the other mentioned contribution to the interaction $\Gamma_{\text{loc}}(\psi^+\phi)(\phi^+\phi)+c.c.$ [third row of Eq. (\ref{Zf}) and second diagram in Fig. \ref{fig:diagramelements1pi}a] is connected to at least one local propagator.
Finally, the determinant $J[\phi^+,\phi]$ provides for the subtraction of diagrams
which are already accounted for in $\Sigma _{\text{loc}}$ and $\Gamma _{\text{loc%
}},$ in particular the bubbles with one (i.e., tadpole terms) and two local
Green's functions, which should be excluded from the diagram technique, see Fig. \ref{fig:diagramelements1pi}b
and Appendix A for details.  

Let us finally comment on the the analytic properties of our new approach: From the diagrammatic elements of the 1PI method in Fig. \ref{fig:diagramelements1pi} one can infer that the situation is completely equivalent to the DF case. For the DF approach, the analyticity of the self-energy has been proven in Ref. \cite{Rubtsov09}. For a complete proof, it is however necessary to show that the corresponding statement holds for the Green's function as well, which remains an open problem. We note that, in practice, no causality violations have been observed in DF and hence we also do not expect violations in our practical calculations.


\section{Ladder approximation in the 1PI approach}

\begin{figure}[t]
 \centering
 \includegraphics[width=0.5\textwidth]{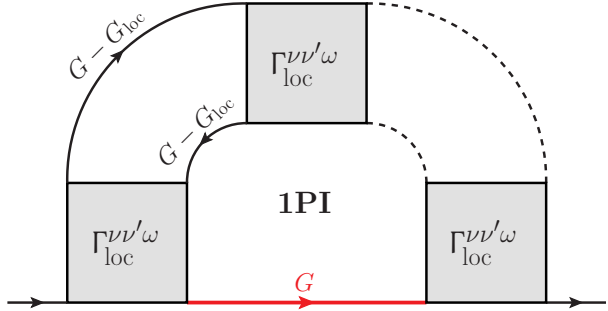}
 \caption{Third order (in terms of the local vertex $\Gamma_{\text{loc},\sigma\sigma'}^{\nu\nu'\omega}$) self-energy diagram
in the 1PI scheme, and ladder extension thereof (indicated by the dashed lines).}
 \label{fig:1PIthirdorder}
\end{figure}

Aiming at a practical application of the 1PI scheme derived in Sec. III, we will now
explicitly consider ladder diagrams for Eq. (\ref{Zf}), see Fig. \ref{fig:1PIthirdorder}. As we mentioned in the introduction,
the restriction to ladder
diagrams is, de facto, the typical approximation scheme adopted in the other diagrammatic extensions of DMFT. Hence, it
represents the natural framework for testing the validity of the 1PI scheme and for comparing its diagrammatic and
physical content against that of DF and D$\Gamma$A.

As for the explicit derivation of the corresponding 1PI expressions for the ladder diagrams, we start from the analysis of all possible bubble-diagrams which can be constructed from the diagrammatic elements for the 1PI approach discussed
in the previous section (see Fig. \ref{fig:diagramelements1pi}).
\begin{figure}[t]
 \centering
 \includegraphics[width=1.0\textwidth]{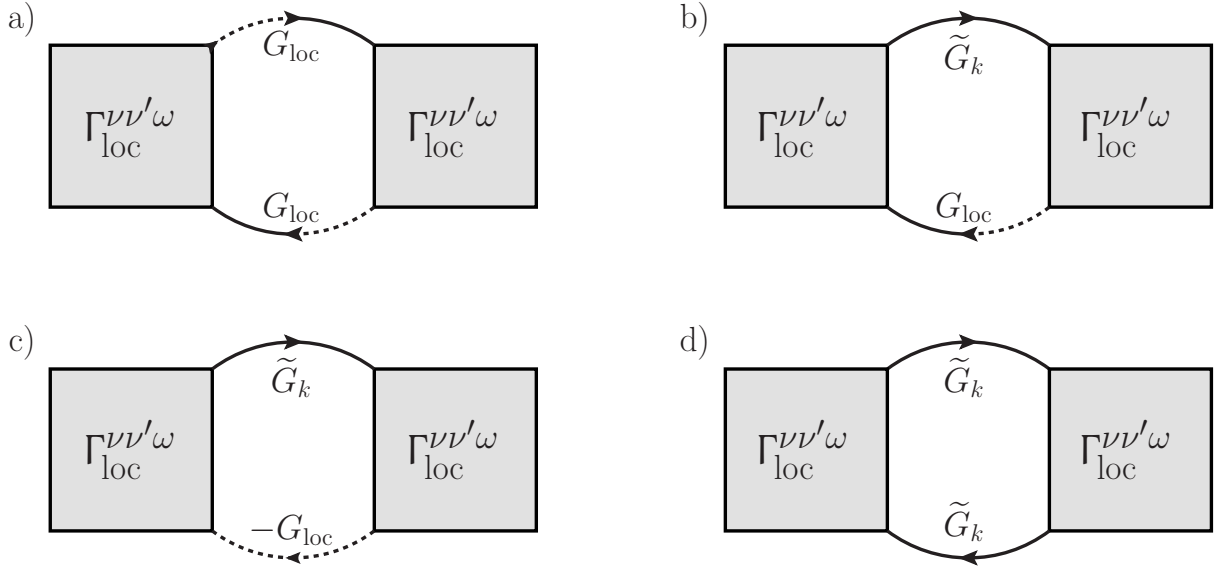}
 \caption{Bubble diagrams for the 1PI approach which can be constructed from the diagrammatic elements shown in Fig. \ref{fig:diagramelements1pi}. Only the diagram d) contributes to the perturbation series.}
 \label{fig:1pibubles}
\end{figure}
Considering all possible bubbles (Fig. \ref{fig:1pibubles}), we observe that the bubble (a) with {\em two} local
Green's functions should not appear in our 1PI corrections to the local self-energy, since it is already included in
$\Gamma_{\rm loc}$ (the contribution of Fig. \ref{fig:1pibubles}a is canceled by the corresponding counterterms
contained in the determinant $J[\phi^+,\phi]$ of the Jacobian, shown by the second diagram of Fig \ref{fig:diagramelements1pi}b). On the other hand, bubble diagrams with a {\em
single}  local Green's function, as depicted in Fig. \ref{fig:1pibubles}b,c vanish due to the fact that
$\widetilde{G}_k$, summed over $\mathbf{k}$, yields zero. Hence,
the ladder part of the diagram for the self-energy can be solely composed of bubbles with {\em two} nonlocal Green's
functions $\widetilde{G}_k$ (see Fig. \ref{fig:1pibubles}d), which makes the considered approach similar to that in Ref.
\cite{noteJanis} with the restriction to the ladder diagrams only. Therefore, the ladder part has to be constructed
solely from
$\Gamma_{\text{loc}}(\phi ^{+}\phi)(\phi ^{+}\phi)$ vertices, except for the leftmost and rightmost vertex which
can be either of the type $\Gamma_{\text{loc}}(\psi^+\phi+\phi^+\psi)(\phi^+\phi)$, connected by {\sl one} local
Green's function $G_{\text{loc}}$, or of the type $\Gamma_{\text{loc}}(\phi^{+}\phi)(\phi ^{+}\phi)$, connected by the
Green's function $\widetilde{G}_k$. 
\begin{figure}[t]
 \centering
 \includegraphics[width=1.0\textwidth]{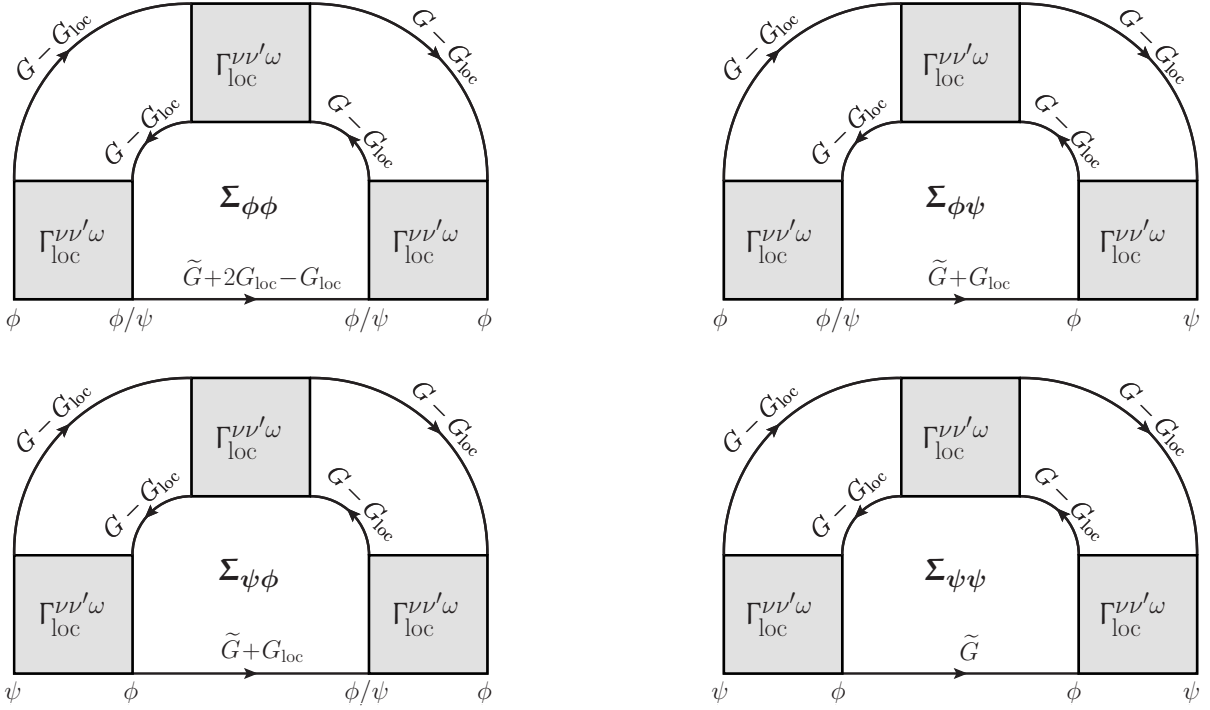}
 \caption{The four components of the matrix $\boldsymbol{\Sigma}$ depicted for diagrams of third order in $\Gamma_{\text{loc}}$.}
 \label{fig:sigmamatrix}
\end{figure}
Hence, as it is illustrated in Fig. \ref{fig:sigmamatrix} for the case of third order (in $\Gamma_{\text{loc}}$) diagrams, the self-energy matrix in the spinor representation,
\begin{align}
\label{sigmamat}
 \boldsymbol{\Sigma}=\begin{pmatrix}
                      \Sigma_{\phi\phi}&\Sigma_{\phi\psi}\\ \Sigma_{\psi\phi}&\Sigma_{\psi\psi}
                     \end{pmatrix},
\end{align}
has only two distinct components:
\begin{align}
 \label{sigmacomponents}
 &\Sigma_{\phi\phi,k}=\Sigma_{\phi\psi,k}=\Sigma_{\psi\phi,k}=\Sigma_{1,k}+\Sigma_{2,k},\nonumber\\
 &\Sigma_{\psi\psi,k}=\Sigma_{1,k},
\end{align}
with $\Sigma_{1,k}$ and $\Sigma_{2,k}$ being defined in the following way:
\begin{equation}
\label{equ:sigma1}
\Sigma _{1,k}\! = \frac{1}{\beta}\sum\limits_{\nu ^{\prime }\nu ^{\prime \prime
}q}\sum\limits_{m=c,s}\!\!A_{m}\Gamma _{\text{loc},m}^{\nu \nu ^{\prime }\omega }\left[
\delta _{\nu ^{\prime }\nu ^{\prime \prime }}\!-\!\widetilde{\chi }^{\nu
^{\prime }}_q \Gamma _{\text{loc},m}^{\nu ^{\prime }\nu ^{\prime \prime
}\omega }\right] _{\nu ^{\prime }\nu ^{\prime \prime }}^{-1}\widetilde{\chi }%
^{\nu ^{\prime \prime }}_q[G_{k+q}\!-\!G_{\text{loc},\nu+\omega}]\Gamma _{\text{loc},m}^{\nu ^{\prime \prime }\nu \omega }\!-\!\Sigma
_{1,k}^{(2)},
\end{equation}
contains the $\widetilde{G}_k$ Green's functions only and
\begin{eqnarray}
\Sigma _{2,k} \!&=&\frac{1}{\beta}\sum\limits_{\nu ^{\prime }\nu ^{\prime \prime
}q}\sum\limits_{m=c,s}\!\!A_{m}\Gamma _{\text{loc},m}^{\nu \nu ^{\prime }\omega }\left[
\delta _{\nu ^{\prime }\nu ^{\prime \prime }}-\widetilde{\chi }^{\nu
^{\prime }}_q \Gamma _{\text{loc},m}^{\nu ^{\prime }\nu ^{\prime \prime
}\omega }\right] _{\nu ^{\prime }\nu ^{\prime \prime }}^{-1}\widetilde{\chi }%
^{\nu ^{\prime \prime }}_q G_{\text{loc},\nu+\omega}\Gamma _{m,%
\text{loc}}^{\nu ^{\prime \prime }\nu \omega }, \label{sigma2}
\end{eqnarray}%
in turn contains the very same ladder but differs by a Green's function $G_{\text{loc}}$ in place of $\widetilde{G}_k$. The contribution $\Sigma _{1,k}^{(2)}=\Sigma ^{(2)}_{\text{d},k} $, which
is the same as the DF second-order diagram in Eq. (\ref{St}), has to be subtracted in Eq. (\ref{equ:sigma1}) to avoid a double counting of the second order diagram (in $\Gamma_{\text{loc}}$) in the ladder series. Note that the matrix inversions in Eqs. (\ref{equ:sigma1}) and (\ref{sigma2}) are performed with respect to the fermionic Matsubara frequencies $\nu'$ and $\nu''$ for each value of $q$ (i.e., for fixed $\omega$ and $\mathbf{q}$).

According to Eqs. (\ref{G_nonint}), (\ref{sigmamat}) and (\ref{sigmacomponents}), the Dyson equation in the spinor formalism reads as:
\begin{equation}
\mathbf{G}^{-1}_k=\boldsymbol{{\cal  G}}_k^{-1}-\boldsymbol{\Sigma}_k=\left(
\begin{array}{cc}
G^{-1}_k-\Sigma _{1,k}-\Sigma _{2,k} & G^{-1}_k-\Sigma _{1,k}-\Sigma _{2,k} \\
G^{-1}_k-\Sigma _{1,k}-\Sigma _{2,k} & G^{-1}_k-G_{\text{loc},\nu}^{-1}-\Sigma _{1,k}
\end{array}
\right).  \label{Sigma_ladd}
\end{equation}
Inverting (\ref{Sigma_ladd}) and performing the summation of the components of the obtained matrix [see Eq. (\ref{equ:gdef})] we obtain the simple result
\begin{equation}
\Sigma_{\text{1PI},k} =\Sigma_{\mathrm{ loc}}(i \nu_n)+\Sigma _{1,k}+\Sigma _{2,k}  \label{Sigma_ladd_res} .
\end{equation}
Expanding the result Eq.\ (\ref{Sigma_ladd_res}) to leading order in
$\widetilde{G}=G-G_{\text{loc}}$, $\Sigma_{2}$ yields zero, while $\Sigma_{1}$ allows to derive Eq.\ (\ref{St}).

From Eq. (\ref{Sigma_ladd_res}) one can see, that the 1PI approach yields no spurious denominator for the
lattice self-energy. Note that in the dual fermion approach \cite{DF}, with the usual restriction to the two-particle local
vertex, only the contribution $\Sigma _{1}$ [with the corresponding denominator, given in the Eq. (\ref{Eq:SDF})] appears, while $\Sigma_{2}$ corresponds to the contributions stemming from
the three-particle  local (one-particle reducible) vertex, see the discussion in Ref. \cite{Katanin12}.

At the same time, both contributions appear on the same ground in the 1PI approach already at the two-particle
 vertex level. As it is shown below, in Sect. V, the contribution $\Sigma _{2,k}$ yields however an enhanced asymptotics of the
self-energy at large frequencies $\nu$. Therefore, at least the high energy part of $\Sigma _{2,k}$ has to be compensated
by the non-ladder diagrams. In this respect, the situation in the 1PI approach is similar
to the ladder approximation within the D$\Gamma$A approximation, where $\lambda$-corrections are needed to obtain the correct asymptotics of the self-energy.

\subsection*{Comparison to the ladder D$\Gamma$A}
To compare the result (\ref{Sigma_ladd_res}) to the ladder D$\Gamma$A, let us represent the reducible local vertex via the irreducible one in a certain particle-hole
channel
\begin{equation}
\Gamma _{\text{ir},s(c)}^{\nu \nu ^{\prime }\omega }=[(\Gamma _{\text{loc},s(c)}^{\nu \nu ^{\prime }\omega })_{\nu \nu ^{\prime }}^{-1}+\chi ^{0,\nu \omega}_%
\mathrm{loc} \delta _{\nu \nu ^{^{\prime }}}]^{-1}  \label{equ:gammairrdef}.
\end{equation}%
We now introduce the vertex
\begin{eqnarray}
\Gamma _{\mathbf{q},s(c)}^{\nu \nu ^{\prime }\omega } &=&[(\Gamma _{\text{ir},s(c)%
}^{\nu \nu ^{\prime }\omega })^{-1}-\overline{\chi}^{\nu}_{q}\delta _{\nu \nu ^{^{\prime }}}]^{-1} , \label{Gamma} \\
\overline{\chi}^{\nu}_q &=&-\frac{1}{\beta}\sum_{\mathbf{k}}G_kG_{k+q}= \chi^{0,\nu\omega}_\text{loc}+\widetilde{\chi}^{\nu}_q  ,\notag
\end{eqnarray}%
where the inversion is performed with respect to the fermionic Matsubara frequencies $\nu$ and $\nu'$. This way, after some algebraic manipulations we obtain
\begin{eqnarray}
\Sigma _{1,k} &=&\frac{1}{\beta}\sum\limits_{\nu ^{\prime }
q}\sum\limits_{m=c,s}A_{m}\Gamma _{\mathbf{q},m}^{\nu \nu ^{\prime }\omega}\;\overline{\chi}^{\nu^{\prime }}_q
\left(G_{k+q}-G_{\text{loc},\nu+\omega}\right)\Gamma
_{\text{ir},m}^{\nu ^{\prime }\nu \omega }-\Sigma _{1,k}^{(2)}, \\
\Sigma _{2,k} &=&\frac{1}{\beta}\sum\limits_{\nu ^{\prime }
q}\sum\limits_{m=c,s}A_{m}(\Gamma _{\mathbf{q},m}^{\nu \nu ^{\prime }\omega }\;\overline{\chi}^{\nu
^{\prime }}_q-\Gamma _{{\rm loc},m}^{\nu \nu ^{\prime }\omega }\chi^{0,\nu^{\prime }\omega}_{\rm loc})G_{\text{%
loc},\nu+\omega}\Gamma _{\text{ir},m}^{\nu ^{\prime
}\nu \omega }.
\end{eqnarray}
In total this yields
\begin{eqnarray}
\Sigma _{\text{1PI},k} &=&\Sigma_{\text{loc},\nu}+\frac{1}{\beta}\sum\limits_{\nu ^{\prime }
q}\sum_{m=c,s}A_{m}\left(\Gamma _{\mathbf{q},m}^{\nu \nu ^{\prime }\omega }\;\overline{\chi}^{\nu
^{\prime }}_q -\Gamma _{{\rm loc},m}^{\nu \nu ^{\prime }\omega }\chi^{0,\nu ^{\prime }\omega}_\text{loc}
\right)\Gamma _{\text{ir},m}^{\nu ^{\prime
}\nu \omega } G_{k+q} -\Sigma _{1,k}^{(2)} \label{Sigma}.
\end{eqnarray}%
This result can be compared to the nonlocal self-energy in D$\Gamma $A as obtained previously in Ref. \cite{DGA},%
\begin{align}
\Sigma_{\text{D}\Gamma\text{A},k}& =\frac{1}{2}{Un}+%
\frac{U}{\beta}\sum\limits_{\nu ^{\prime }q}\overline{\chi}^{\nu ^{\prime }}_q\left( A_s \Gamma _{\mathbf{q},s}^{\nu \nu ^{\prime }\omega
}-A_c \Gamma _{\mathbf{q},c}^{\nu \nu ^{\prime }\omega }\right.
\left. +\frac{1}{2}\Gamma _{\text{loc},c}^{\nu \nu ^{\prime }\omega }-\frac{1}{2}\Gamma _{\text{loc},s}^{\nu \nu ^{\prime }\omega }\right) G_{k+q}.
\label{Eq:final}
\end{align}
From the comparison of the above expression to the 1PI ladder self-energy, Eq. (\ref{Sigma}), we can recognize an important difference: the bare interaction $U$ in
Eq. (\ref{Eq:final}) is replaced by the local particle-hole irreducible vertex $\Gamma _{\text{ir}}$ in Eq. (\ref{Sigma}), which is discussed diagrammatically in the next subsection.

\subsection*{Differences in the 1PI, DF and D$\Gamma$A diagrammatics}
\begin{figure}[t]
 \centering
 \includegraphics[width=1.0\textwidth]{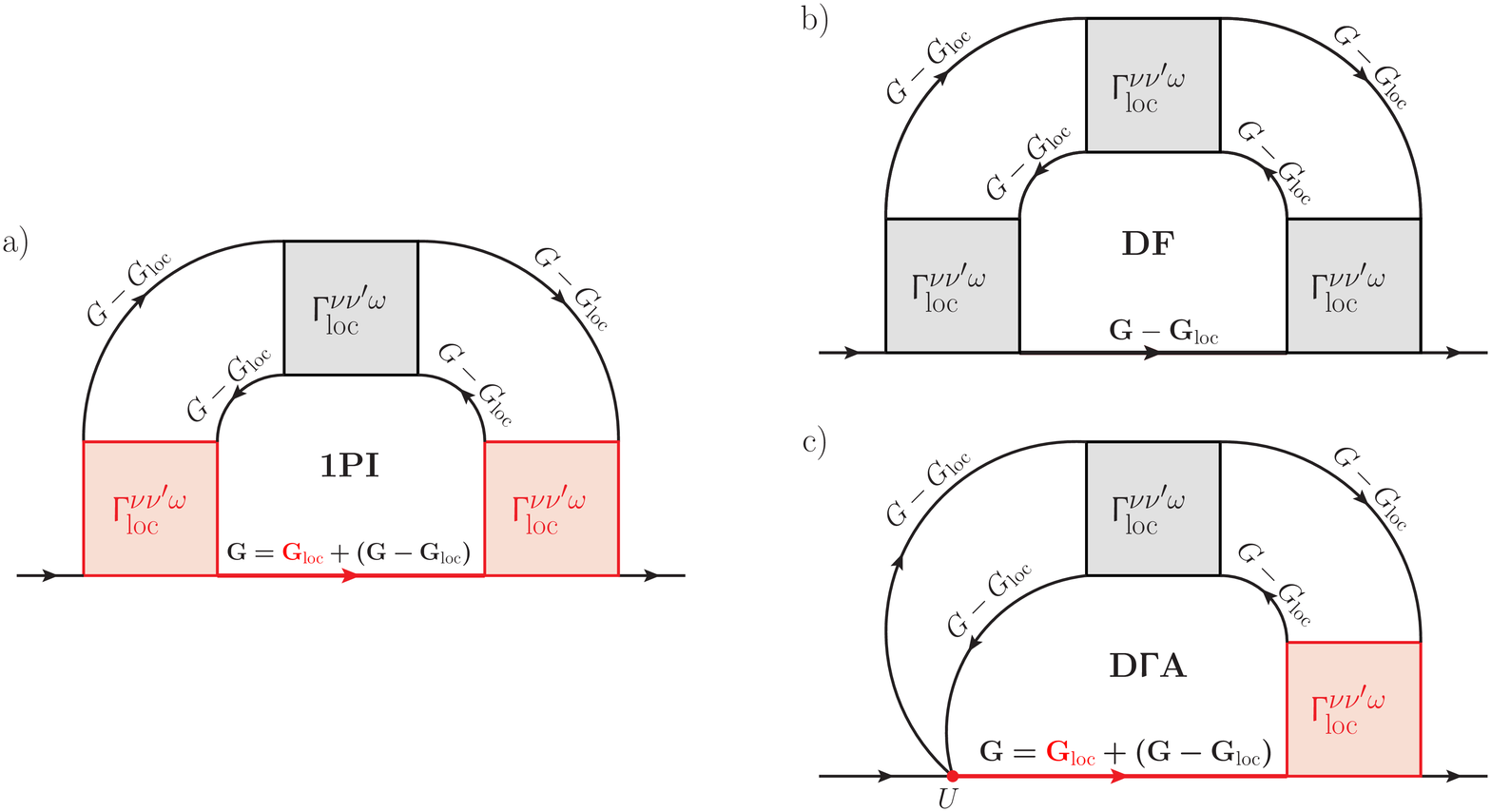}
 \caption{Third order (in terms of the local vertex $\Gamma_{\text{loc},\sigma\sigma'}^{\nu\nu'\omega}$) diagrams for 1PI (a), DF (b) and its corresponding  D$\Gamma$A counter part (c). The contribution of (a part of) the one-particle reducible three-particle  vertex is marked in red in the 1PI and D$\Gamma$A diagrams.}
 \label{fig:dfdgathirdorder}
\end{figure}

The different diagrammatic content of the ladder 1PI, ladder DF and ladder D$\Gamma$A approaches is readily individuated by a direct inspection of the corresponding diagrams. We will start by considering a typical third-order diagram of the 1PI ladder series, shown in Fig. \ref{fig:dfdgathirdorder}a. Comparing to the corresponding diagram of the DF approach (Fig. \ref{fig:dfdgathirdorder}b), it is evident that the latter does not include the term where the fermionic line at the bottom (bold red line in Fig. \ref{fig:dfdgathirdorder}a) corresponds to a local Green's function. This is due to the fact that in the dual fermion space the propagation occurs via purely nonlocal Green's functions $G-G_{\mathrm{loc}}$.
Hence, when only the two-particle local vertex is considered as interaction among the dual fermions, there is no way to generate local Green's functions in the DF ladder diagrams. The difference between the diagrams of Fig. \ref{fig:dfdgathirdorder}a and Fig. \ref{fig:dfdgathirdorder}b corresponds to the contribution of the three-particle vertex in the DF approach (red part in Fig. \ref{fig:dfdgathirdorder}a).

As in 1PI and in contrast to DF, also the corresponding D$\Gamma$A diagram (Fig.\ \ref{fig:dfdgathirdorder}c) contains
the full Green's function $G = G_{\mathrm{loc}} + (G - G_{\mathrm{loc}})$, which also yields ``mixed'' terms with $G - G_{\text{loc}}$ propagators in the ladder part of the diagram and one local $G_{\mathrm{loc}}$ outside the ladder (bottom of the diagram). Again, as for the 1PI diagram, the part of Fig. \ref{fig:dfdgathirdorder}c colored in red corresponds to the contribution of the three-particle  vertex in the DF approach.

At the same time, one should emphasize that the D$\Gamma$A ladder diagrams, as those depicted in Fig. \ref{fig:dfdgathirdorder}c, evidently represent only a subset of the 1PI ladder diagrams. This can be easily understood from a comparison of Fig. {\ref{fig:dfdgathirdorder}a and Fig. \ref{fig:dfdgathirdorder}c: In the 1PI approach all vertices appearing in the diagrams are the dynamical ones ($\Gamma_{\text{loc}}$), while in D$\Gamma$A one of the vertex functions is replaced by its lowest-order counterpart, i.e., the bare interaction $U$. On the other hand, the 1PI ladder diagrams themselves are in turn just a subset of the more general set of diagrams generated by employing the parquet equations for the D$\Gamma$A instead of the ladder approximation.

What does the formal difference between 1PI and D$\Gamma$A mean physically? As it is illustrated in Fig.\ref{fig:1pitodga}, the extra diagrams of 1PI  correspond to considering {\sl nonlocal} corrections to the irreducible vertex in the selected channel [Eq. (\ref{equ:gammairrdef})], while in ladder D$\Gamma$A calculations perfect locality of this vertex is assumed. Obviously the assumption of locality of the irreducible spin- and charge-vertex does not hold for the full D$\Gamma$A where nonlocal corrections to these vertices are generated via the self-consistent solution of the parquet equations. Hence, while, in general, the inclusion of a larger number of diagrams does not guarantee an improvement of a given approximation, in our case the additional nonlocal corrections for the irreducible (spin- and charge-) vertices are physically justifiable through the comparison with the full (parquet) D$\Gamma$A approach.

In order to demonstrate the differences between 1PI and D$\Gamma$A also analytically in the most transparent way, we
can expand the D$\Gamma$A ladder expression for the self-energy [Eq. (\ref{Eq:final})] by representing
$\overline{\chi} _{q}^{\nu ^{\prime }}$ as a sum of local and nonlocal parts, $\chi ^{0,\nu^{\prime }\omega }_\text{loc}%
+\widetilde{\chi }_{q}^{\nu ^{\prime }}$. Expanding to first order in $\widetilde{\chi
}_{q}^{\nu ^{\prime }}$, we obtain%
\begin{eqnarray}
\label{SigmaFin}
\Sigma^{(2)}_{\text{D}\Gamma \text{A},k}=\Sigma _{\text{loc},\nu}+\frac{1}{\beta}\sum\limits_{\nu ^{\prime }\nu''q}\left[%
A_s\Gamma _{\text{loc},s}^{\nu \nu ^{\prime \prime }\omega }\widetilde{\chi }%
_{q}^{\nu ^{\prime \prime }}\left(\overline{\Gamma }_{\text{loc},s%
}^{\nu ^{\prime \prime }\nu ^{\prime }\omega }-\frac{U}{2}\delta_{\nu'\nu''}\right)\right. \notag \\
\left. +A_c\Gamma _{\text{loc},c}^{\nu \nu ^{\prime \prime }\omega }%
\widetilde{\chi }_{q}^{\nu ^{\prime \prime }}\left(\overline{%
\Gamma }_{\text{loc},c}^{\nu ^{\prime \prime }\nu ^{\prime }\omega }+\frac{U}{2}\delta_{\nu'\nu''}\right)%
\right]
\widetilde{G}_{k+q},
\end{eqnarray}
where
\begin{align}
\overline{\Gamma }_{\text{loc},s(c)}^{\nu \nu ^{\prime }\omega }&
=\pm U\left[\delta_{\nu\nu'}-\Gamma _{\text{ir},s(c)}^{\nu \nu ^{\prime}\omega }\chi
_{0\omega ,\text{loc}}^{\nu}\right]^{-1} \\ \nonumber
& =\pm U\sum_{\nu''}\Gamma _{\text{loc},s(c),}^{\nu \nu ^{\prime \prime }\omega }\left[\Gamma
_{\text{ir},s(c)}^{\omega }\right]_{\nu ^{\prime \prime }\nu ^{\prime }}^{-1}.\label{equ:defgammabar}
\end{align}
Expanding the corresponding expression for the 1PI self-energy in Eq. (\ref{Sigma}) in a similar manner, one obtains
$\Sigma ^{(2)}_{\text{1PI},k}= \Sigma _{\text{loc},\nu%
}+ \Sigma ^{(2)}_{\text{d},k}$.
Comparing this result to the corresponding D$\Gamma$A one [Eq. (\ref{SigmaFin})] one observes two differences: (i) The
factor $1/2$ in Eq. (\ref{St}), which avoids double counting of diagrams is replaced by an explicit subtraction of
double counting terms $\pm U/2$ in Eq. (\ref{SigmaFin}) for the D$\Gamma$A. The reason for this is the ``asymmetric''
form of the D$\Gamma$A self-energy correction compared to the 1PI one (bare $U$ in D$\Gamma$A vs. the full vertex on in
1PI on the left hand side of the self-energy diagrams, see Fig. \ref{fig:dfdgathirdorder}. (ii) The second, more
important, difference between the two expressions is that $\overline{\Gamma }_{\text{loc},s(c)}^{\nu \nu ^{\prime
}\omega }$ in Eq. (\ref{SigmaFin}) is replaced by $\Gamma_{\text{loc},s(c)}^{\nu \nu ^{\prime }\omega }$ in 
$\Sigma ^{(2)}_{\text{1PI}}$. Hence, the difference between $\overline{\Gamma }_{\text{loc},s(c)}^{\nu \nu ^{\prime }\omega }$ and
$\Gamma_{\text{loc},s(c)}^{\nu \nu ^{\prime }\omega }$ marks a particular set of nonlocal corrections to the
self-energy, naturally generated in the 1PI ladder diagrams, but neglected in the ladder expansions of the D$\Gamma $A.

\begin{figure}[t]
 \centering
 \includegraphics[width=0.7\textwidth,angle=270]{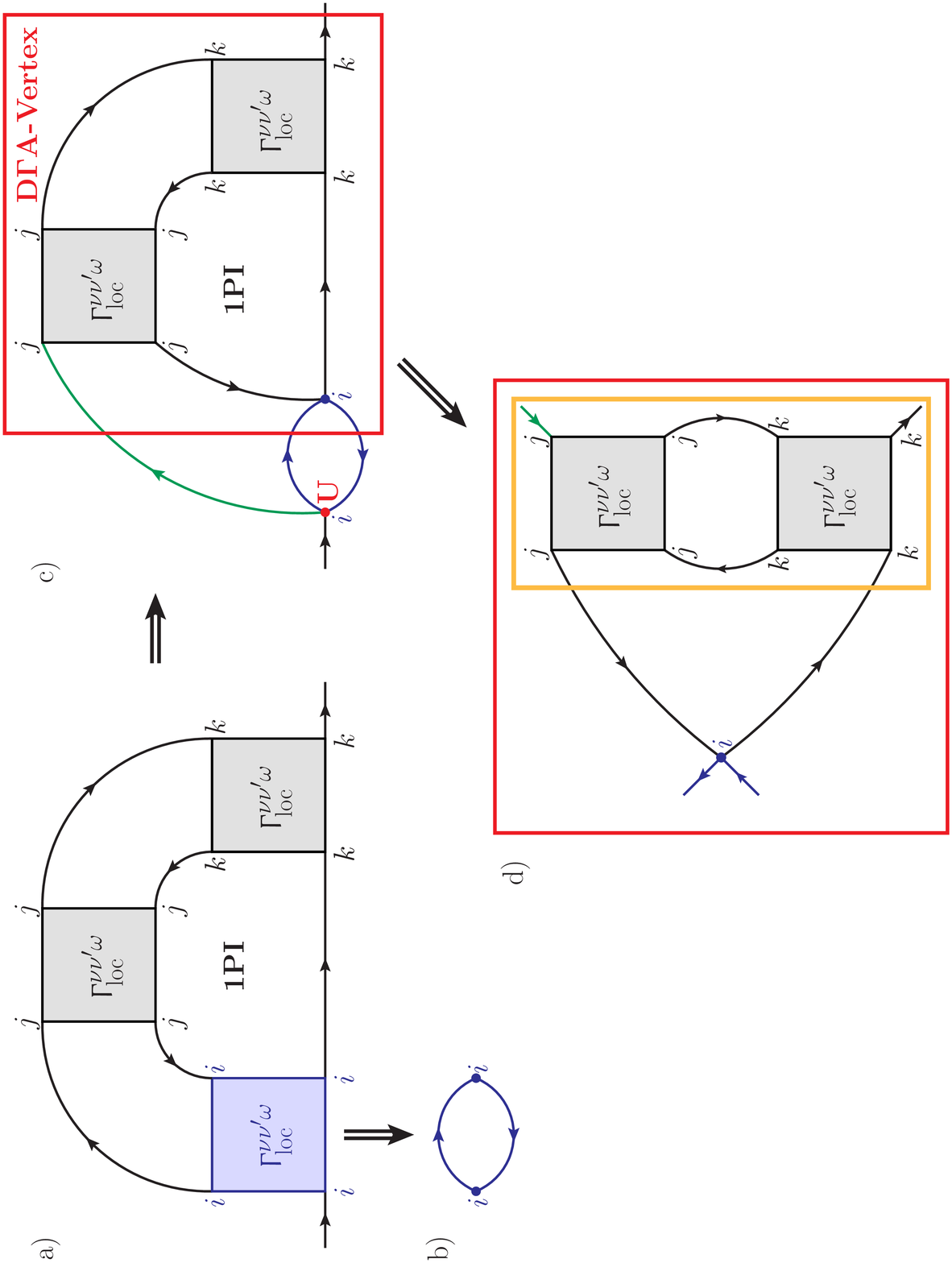}
 \caption{Expressing a 1PI diagram in terms of D$\Gamma$A where there is only a bare $U$ at the left-hand side of the diagram: We start from the specific 1PI diagram a) and consider the specific contribution to the leftmost vertex shown in b), so that a) becomes c). D$\Gamma$A, on the other hand, requires a bare $U$ on the leftmost side (see Fig. \ref{fig:dfdgathirdorder}c). Hence, the entire red box has to be interpreted as a D$\Gamma$A generated reducible vertex. The red box can be deformed to d). The yellow box in diagram d) clearly contains nonlocal contributions to the vertex irreducible in the longitudinal channel. To generate these in the D$\Gamma$A the full parquet treatment would be necessary.}
 \label{fig:1pitodga}
\end{figure}

The interpretation of the ladder 1PI expression derived in this section can be summarized as follows: in the ladder approximation, the 1PI diagrams include terms not present in D$\Gamma$A and DF. In the latter approaches, these are generated by going {\sl beyond} the ladder approximation to D$\Gamma$A and beyond the two-particle vertex in DF, respectively. The numerical effort of performing a ladder 1PI calculation is much smaller compared to the full (parquet-based) D$\Gamma$A, or to the DF with the three-particle  vertex. In a sense the 1PI approach better utilizes the information contained in the single-particle Green's function and two-particle vertex.

\section{Numerical results}
\label{Sec:results}

In this section, we present numerical results for nonlocal corrections to the self-energy of the two-dimensional
Hubbard model obtained by means of the ladder 1PI formalism and compare them with the corresponding DF and D$\Gamma$A results.
We consider the relevant case of the Hubbard model on a (two-dimensional) square lattice with nearest-neighbor hopping $t$ at half-filling, where the effect of nonlocal correlations beyond DMFT is expected to be particularly strong. Note that in the following all energy scales, such as the Hubbard interaction parameter $U$ and the temperature $T=1/\beta$, will be given in units of the half bandwidth $W/2=4t=1$. Furthermore, one should bear in mind that for the half-filled Hubbard model the self-energy evaluated for $\mathbf{k}$-points at
the Fermi-surface is purely imaginary as a function of Matsubara frequencies (besides the constant Hartree-contribution $\frac{Un}{2}$). Hence, in order to
keep the notation as simple as possible, $\Sigma$ refers to the {\sl imaginary part} of the self-energy,
i.e., $\Sigma\mathrel{\widehat{=}}\text{Im}\Sigma$, in the sections below.

Before presenting our numerical results in the next two subsections, let us stress that the only possibility to perform
a one-by-one comparison between the diagrammatic methods stands for the (non-self-consistent) one-shot calculations. As
discussed in Sec. IV, only in this case the exact relations between the three different approaches and their
diagrammatic content can be identified. Hence, this analysis is performed first. The obtained results do not
necessarily represent the final, physical results of the three methods. In a separate subsection, we therefore look at
the trends emerging when going beyond the one-shot calculations. We note that because of the different ways the
self-consistency is implemented (inner and outer self-consistency loop in DF \cite{note_scDF}, Moriyaesque
$\lambda$-correction \cite{note_scDGA} in D$\Gamma$A and 1PI), as well as the different possible levels of
approximation (ladder- or parquet-diagrams) an identification of equivalent levels of approximation as in the one-shot
case is not possible. Also for keeping the comparison among
different methods as precise as possible, we present our numerical results on the Matsubara frequency axis only,
avoiding the additional, and  to some extent uncontrolled, effects of an analytic continuation.

\subsection{One-shot calculations}

In this subsection, we will focus on non-self-consistent one-shot calculations for nonlocal corrections to the (local)
DMFT self-energy: this approach represents an expansion around DMFT, where the auxiliary local AIM [Eq. (\ref{LDMFT})]
is not changed w.r.t. DMFT and the DMFT Green's functions [Eq. (\ref{Glattice})] are not renormalized by a feedback of
the nonlocal self-energy. As one can understand from the discussion in the previous sections, examining
(non-self-consistent) one-shot calculations corresponds to considering well-defined sets of diagrams for the lattice
electrons. This way we are able to individuate the general trends obtained by the three approaches (1PI, DF and
D$\Gamma$A) emerging purely from their different diagrammatic content.

 For the sake of conciseness, we will mainly
discuss the numerical results obtained with ladder calculations, since they are most frequently adopted in previous
papers \cite{LDFA,DGA2,GangLi2008,DGA3}, and the inclusion of ladder diagrams proved to be essential to correctly
describe crucial features of the two- and three dimensional physics. Examples are the pseudogap \cite{LDFA} in $d=2$ or
the critical exponents in $d=3$ dimensions \cite{DGA3}.

\begin{figure}[t]
\centering
 \includegraphics[width=0.9\textwidth]{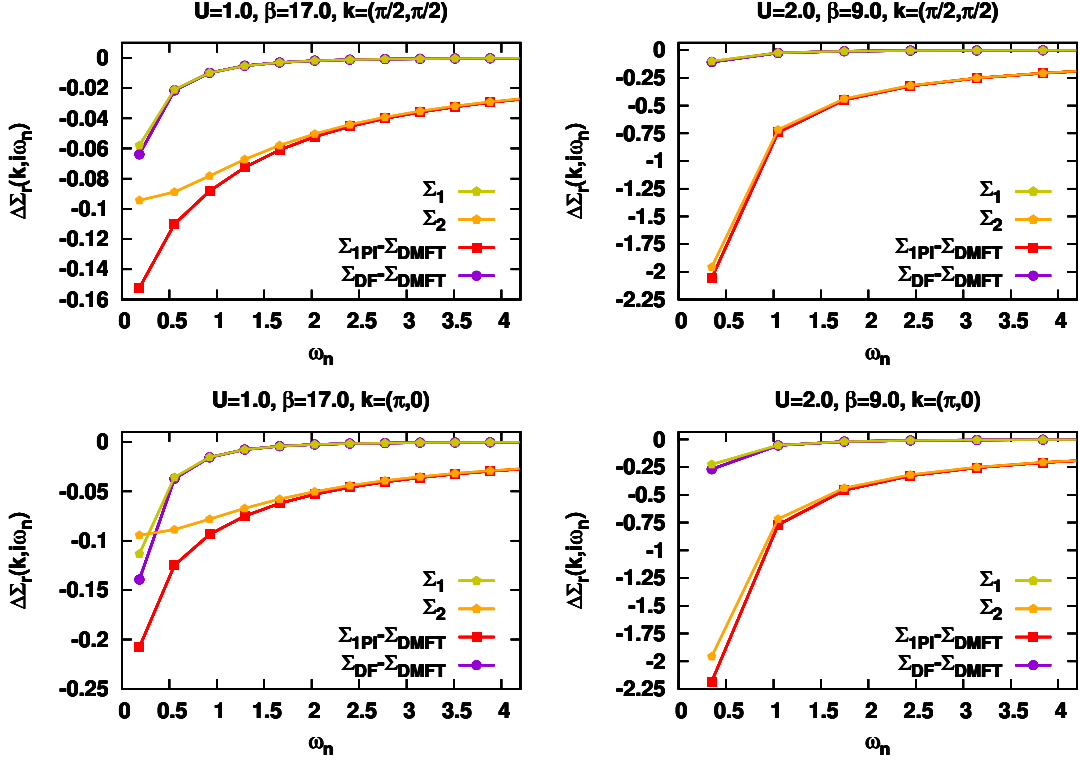}
 \caption{Nonlocal corrections $\Delta\Sigma_{\text{r}}(\mathbf{k},i\omega_n)\!=\!\Sigma_{\text{r}}(\mathbf{k},i\omega_n)\!-\!\Sigma_{\text{loc}}(i\omega_n)$ (r$=$1PI [Eq. (\ref{Sigma_ladd_res})] and DF [Eq. (\ref{Eq:SDF})], respectively) to the DMFT (local) self-energy for
   the $d=2$ Hubbard model on a square-lattice
 at half filling for two different values of $U$,  two different {\bf
   k}-points on the Fermi surface (i.e., {\bf k}
 =$(\frac{\pi}{2},\frac{\pi}{2})$, nodal point, and {\bf k}
 =$(\pi,0)$, anti-nodal point), and temperatures
 slightly above  the corresponding $T_{\text{N}}$ of DMFT. For the 1PI results the single contributions $\Sigma_1$ [Eq. (\ref{equ:sigma1})] and $\Sigma_2$ [Eq. (\ref{sigma2})] are also shown separately. Note, all self energies are purely imaginary; this imaginary part is shown.}
 \label{fig:1pivsdf}
\end{figure}

In Fig. \ref{fig:1pivsdf} we present our results for one-shot calculations of the nonlocal corrections to the DMFT self-energy, $\Delta\Sigma_{\text{r}}(\mathbf{k},i\omega_n)\!=\!\Sigma_{\text{r}}(\mathbf{k},i\omega_n)\!-\!\Sigma_{\text{loc}}(i\omega_n)$ for r$=$1PI [Eq. (\ref{Sigma_ladd_res})] and DF [Eq. (\ref{Eq:SDF})], respectively, on the Matsubara frequency axis. For the 1PI approach we also show its two contributions $\Sigma_1$ [Eq. (\ref{equ:sigma1})] and $\Sigma_2$ [Eq. (\ref{sigma2})] separately. Note, that since no self-consistent adaption of the underlying local model is performed, the local self-energy coincides with the DMFT one, i.e., $\Sigma_{\text{loc}}(i\omega_n)\!=\!\Sigma_{\text{loc}}^{\text{DMFT}}(i\omega_n)$. Data for weak- ($U=1$) and intermediate coupling ($U=2$) and for two different {\bf k}-points on the Fermi surface are presented. The temperature has been chosen to be slightly above the onset of the antiferromagnetic ordering (N\'eel temperature, $T_{\rm N}^{\rm DMFT}$) obtained in DMFT, aiming to maximize the effect of nonlocal correlations.
One can see that, quite generally, the nonlocal corrections in the considered approaches increase the imaginary part of the self-energy,
making its low-frequency dependence less metallic.
Comparing the relative magnitudes of the nonlocal corrections shown in Fig. \ref{fig:1pivsdf}, the
contribution of $\Sigma_1$ of the 1PI approach appears always rather small even though the $U$ and $T$ values have been selected very close to the antiferromagnetic instability of DMFT.
The reason for this behavior is that in $\Sigma_1$ one has to perform {\bf k}-summations over terms containing $G-G_{\text{loc}}$, which yields small results since in a one-shot calculation, $\sum_{\bf k} G_k - G_{\text{loc}}=0$  because of the DMFT self consistency  [Eq.\ (\ref{sc})]. Let us also note that in one-shot calculations, the $\Sigma_1$-part of the 1PI correction [Eq. (\ref{equ:sigma1})] almost exactly coincides with the DF correction $\Sigma_{\rm DF}-\Sigma_{\text{DMFT}}$, albeit without the denominator in Eq. \eqref{Eq:SDF}. For the data presented here, the effect of the denominator is found to be rather small.
On the contrary, in $\Sigma_2$ a mixing of local and nonlocal contributions occurs, because one single Green's function $G_{\text{loc}}$ enters instead of $G-G_{\text{loc}}$ [see Eq. (\ref{sigma2})]. Hence this term becomes significantly larger than $\Sigma_1$.

However, as it was already mentioned in Sect. IV, the contribution $\Sigma_2$ displays an enhanced high-frequency asymptotics,
while $\Sigma_1$ decays faster than $\frac{1}{i\omega_n}$ and preserves the exact asymptotic behavior of  the self-energy
when added to the local self-energy of DMFT. The reason
for this is again that $\Sigma_1$  is constructed from $G-G_{\text{loc}}$ only, which decays as
$\frac{1}{(i\omega_n)^2}$. $\Sigma_2$ has an explicit $\frac{1}{i\omega_n}$ contribution from the
$G_{\text{loc}}$-term, which leads to a (spurious) correction of the already exact  $\frac{1}{i\omega_n}$ behavior of
the DMFT self-energy. We note here that the enhanced asymptotic of $\Sigma_2$ and, hence, of the 1PI approach, is
exactly the same as in  D$\Gamma$A\cite{phdrorhinger} as one can observe in Fig. \ref{fig:1pivsdga}.
Similarly to the D$\Gamma$A case, the enhanced asymptotic is corrected either by treating the full
parquet set of diagrams, or enforcing the condition $\sum_{\bf
q} \chi({\bf q}) = \chi_{\text{AIM}} $ at the ladder level via  Moriyaesque $\lambda$-corrections \cite{DGA2}, see the
results in the next subsection.

 In Fig. \ref{fig:1pivsdga} we plot the self-energy obtained from one-shot ladder calculations for 1PI, DF and D$\Gamma$A in comparison to DMFT. For 1PI and D$\Gamma$A, nonlocal corrections are large as expected from the proximity to the DMFT N\'eel temperature.
In the weak-coupling regime (i.e., for $U=1.0$), one further observes that the 1PI correction is smaller than the corresponding D$\Gamma$A correction. The reason for this is that the $U$ appearing in the D$\Gamma$A equation (\ref{Eq:final}) is replaced by the irreducible vertex in the 1PI formula. At small values of the interaction parameter $U$, the (irreducible) vertex is smaller \cite{DGA,rohringer2012} than the bare interaction due to metallic screening. Therefore, nonlocal corrections obtained within the 1PI formalism tend to be smaller than the one obtained in D$\Gamma$A.

The situation is completely reversed in the strong coupling regime ($U=2$). Here, the local (irreducible) vertex is strongly enhanced\cite{DGA,rohringer2012,schaefer2013} compared to the bare Hubbard interaction $U$, due to the formation of the local moment in the proximity of the Mott phase. Hence, the 1PI self-energy correction is larger than that obtained in D$\Gamma$A.

\begin{figure}[t]
 \centering
 \includegraphics[width=0.9\textwidth]{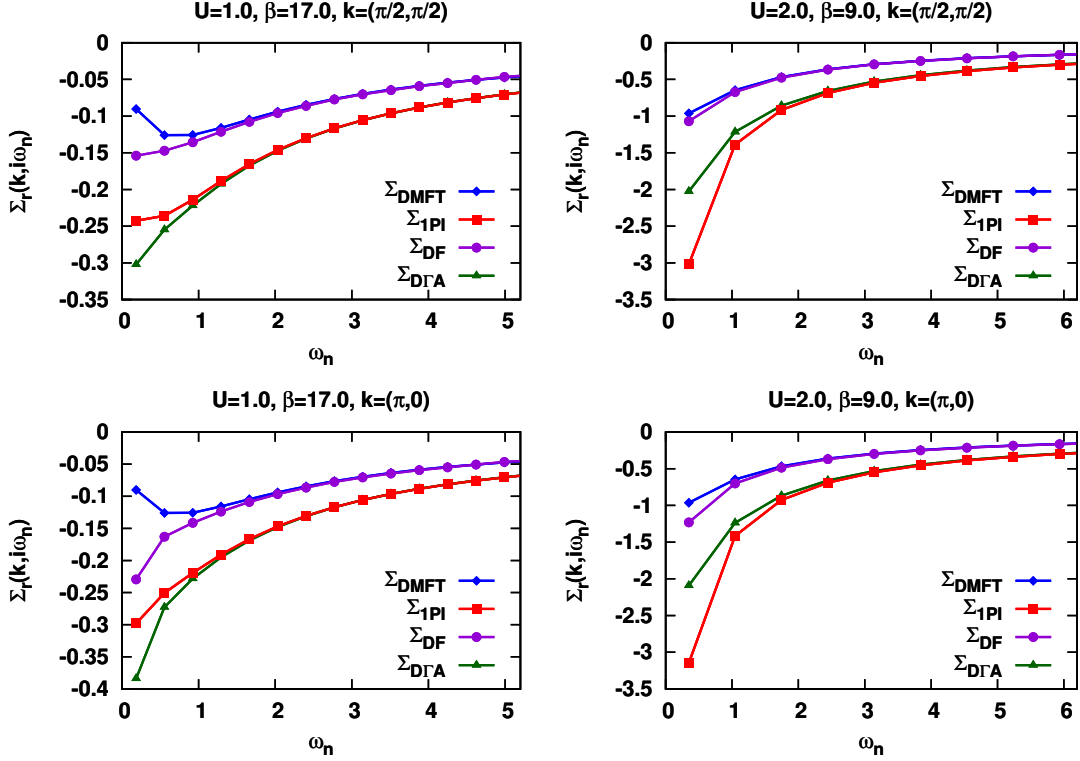}
 \caption{Self-energies (imaginary part) obtained with one-shot calculations for the 1PI approach vs. DF, D$\Gamma$A and DMFT self-energies for the same parameters as in Fig. \ref{fig:1pivsdf}.}
 \label{fig:1pivsdga}
\end{figure}

In the present implementation of 1PI and D$\Gamma$A the calculation of the Neel-temperature $T_{\text{N}}$ by means of a $\lambda$-correction
is purely based on the asymptotic behavior of the (nonlocal) self-energy. This is the same in both approaches and, hence, one would get the same transition temperatures. However, an improved scheme of $\lambda$-corrections or a self-consistent treatment of these theories is expected to yield different $T_{\text{N}}$'s.  In Ref.~\cite{DGA3}
 $T_{\rm N}$ was found smaller in  D$\Gamma$A than the one estimated in DCA \cite{DCAdata} or in lattice quantum Monte Carlo \cite{QMCdata} at weak-coupling,  indicating a possible overestimation of  the nonlocal correlation effects. As it was argued in Ref.~\cite{DGA3}, nonlocal corrections to the charge- and particle-particle irreducible channels, which can be rigorously included only by performing the D$\Gamma$A at the parquet level, might be responsible for this. Hence, the 1PI approach, which partly takes such corrections into account (see Fig.\ \ref{fig:1pitodga}), is rather promising to
improve the agreement between the diagrammatic and the cluster estimations of $T_{\rm N}$ in the Hubbard model, even in the (self-consistent) ladder approximation. This may also hold true in the strong-coupling regime, where $T_{\rm N}$ was slightly larger in ladder-D$\Gamma$A than in the cluster methods.

As for the comparison with the DF self-energy one can see that it is smaller than the corresponding 1PI and D$\Gamma$A ones. The reason for this is the same as discussed for the contribution $\Sigma_1$ to the 1PI self-energy. However, one should consider, that the different ways of self-consistency for 1PI, DF and D$\Gamma$A can change this situation dramatically.

\subsection{Self-consistency and Moriyaesque $\lambda$-corrections}

The analysis of the one-shot results has shown the existence of a well-defined hierarchy in the relative magnitude of the nonlocal corrections. It is however expected that the overall size of the nonlocal corrections will be strongly modified by the inner and outer self-consistency loops in DF \cite{LDFA} on the one hand and the inclusion of the Moriyaesque $\lambda$-corrections in D$\Gamma$A \cite{DGA2} and 1PI \cite{1pi_selfconsistency} on the other.
These effects are briefly analyzed in this subsection.

\begin{figure}[t]
 \centering
 \includegraphics[width=0.9\textwidth]{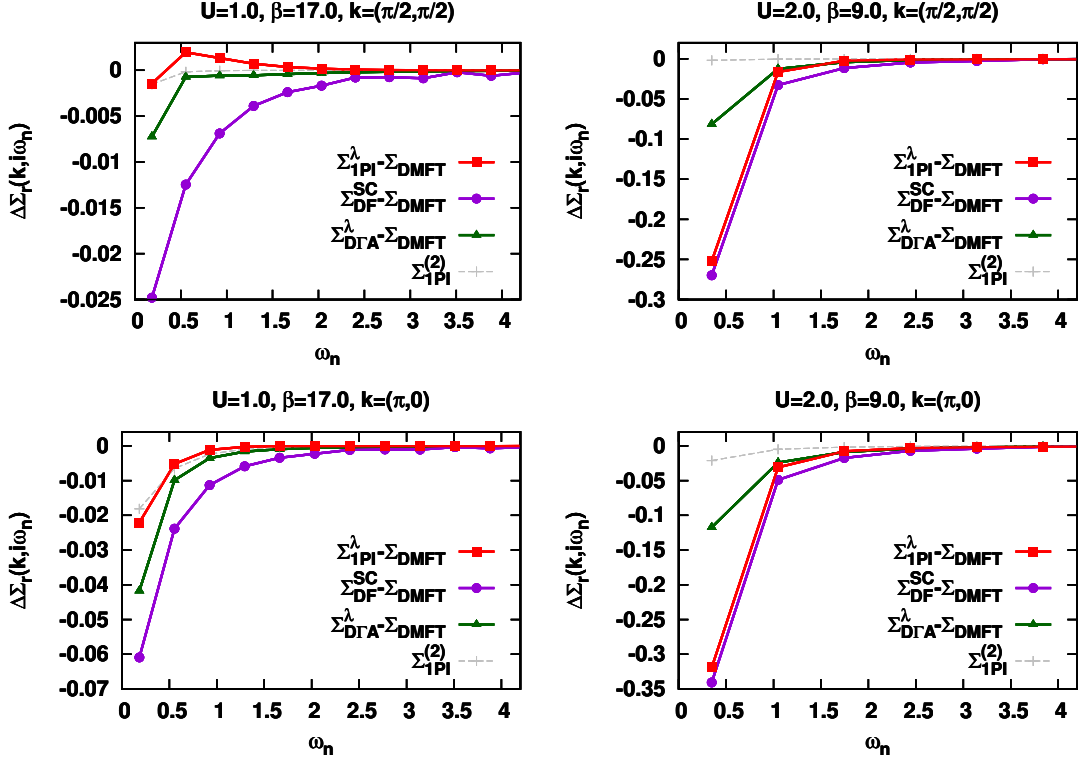}\\
 \caption{Nonlocal corrections $\Delta\Sigma_{\text{r}}(\mathbf{k},i\omega_n)\!=\!\Sigma_{\text{r}}(\mathbf{k},i\omega_n)\!-\!\Sigma_{\text{loc}}(i\omega_n)$ (r$=$1PI, DF and D$\Gamma$A, respectively) as in Fig. \ref{fig:1pivsdf}, but for calculations including Moriyaesque $\lambda$-corrections (1PI and D$\Gamma$A) or self-consistency (DF). Besides the ladder 1PI results we also include the results from the second-order diagram $\Sigma_{\text{1PI}}^{(2)}(\mathbf{k},i\omega_n)=\Sigma_{\text{d}}^{(2)}(\mathbf{k},i\omega_n)$ given in Eq. (\ref{St}).}
 \label{fig:lambdacorronly}
\end{figure}

The results of the self-consistent DF, D$\Gamma$A, and 1PI approaches are presented in Fig. \ref{fig:lambdacorronly}.
Comparing them to Fig. \ref{fig:1pivsdf}, one observes that the inclusion of the $\lambda$-corrections in D$\Gamma$A
and 1PI (which reduces the value of $T_{\rm N}$ from the overestimated DMFT value
) leads to a significant reduction of the nonlocal corrections to the self-energy (note the different
scales in the two figures). This has been observed previously for D$\Gamma$A \cite{DGA2,DGA3}. Hence, the
$\lambda$-corrected results become  much more similar to those obtained in {\sl self-consistent} DF calculations. In
particular, at strong coupling, 1PI and DF agree rather well.
The previously mentioned
hierarchy in the relative magnitude of the nonlocal corrections to DMFT of 1PI and D$\Gamma$A results is fully preserved
by the Moriyaesque $\lambda$-corrections (see Fig. \ref{fig:lambdacorronly}):
\begin{figure}[t]
 \centering
 \includegraphics[width=0.9\textwidth]{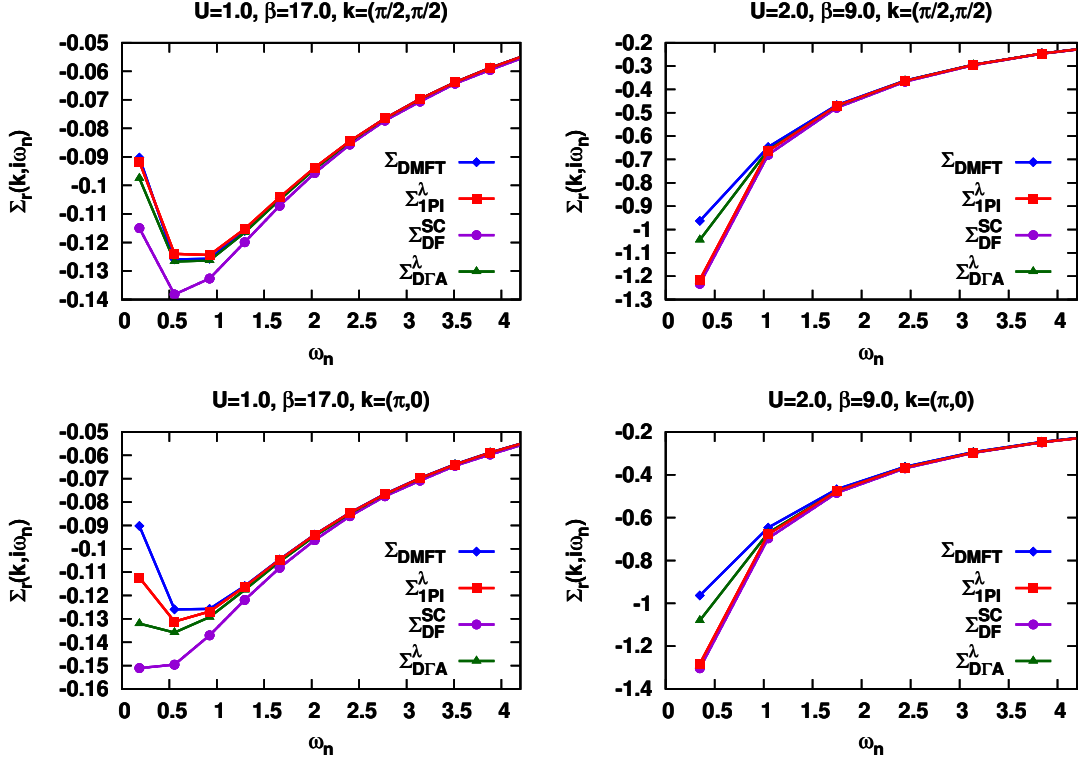}
 \caption{Self-energies obtained with the 1PI approach including $\lambda$-corrections vs. self-consistent DF, $\lambda$-corrected D$\Gamma$A and DMFT self-energies for the same parameters as in Fig. \ref{fig:lambdacorronly}.}
 \label{fig:1PIvsDGAscsigma}
\end{figure}
At weak coupling ($U=1.0$) the 1PI corrections remain smaller than the  D$\Gamma$A ones  due to the metallic screening
of the irreducible vertex, while in the strong coupling regime ($U=2.0$) the enhancement of the same vertex due to the
vicinity of the MIT leads to larger corrections for the 1PI approach with respect to the D$\Gamma$A.
Note that the small value of the nonlocal part of the self-energy in the 1PI approach at $U=1.0$
(especially in the nodal direction) may result from a simplified way of considering self-consistent effects
through the $\lambda$-correction. Since this correction is determined solely from the asymptotic behavior of
the self-energy at large frequencies, it may yield an overestimation of the effect of non-ladder diagrams in the 1PI
approach in the low-frequency region.

In Fig. \ref{fig:1PIvsDGAscsigma} we present the corresponding results for the self-energies.
For $U=1.0$ one can see, that at the considered temperature one observes metallic behavior in all the
approaches, except for the DF data in the $(\pi,0)$ direction. We have verified, however, that even for
this relatively small value of $U$ the nonlocal 1PI corrections, though smaller than the D$\Gamma$A and DF ones, eventually
overcome the metallic behavior of the DMFT self-energy at sufficiently small temperatures, consistent with the results of Ref.\ \cite{bluemer}. 
We emphasize that, for $U=1.0$, cluster extensions of DMFT would predict, instead, a low-temperature metallic phase\cite{CDMFT_MIT}. This confirms the necessity of including long-range antiferromagnetic fluctuations beyond DMFT
in order to capture correctly the interplay of the Mott-Hubbard 
transition and antiferromagnetism (at $T=0$), whose nature gradually changes from Slater to Heisenberg\cite{AF,AF_DMFT}.

\section{Conclusions}

In this paper, we have developed a new one-particle irreducible (1PI) approach for including nonlocal spatial correlations on top of the local correlations of dynamical mean-field theory. We have compared it with the existing state-of-the-art diagrammatic extensions of DMFT, namely dual fermion (DF) and dynamical vertex approximation (D$\Gamma$A).

Starting point of the 1PI approach is the generating functional formalism in the functional integral representation.
Similar as in the DF theory, we decouple local and nonlocal degrees of freedom by means of a Hubbard-Stratonovich
transformation and integrate out the local degrees of freedom. However, instead of expanding the logarithm of the local
generating functional in the source fields, which would lead to local one-particle reducible two- and more-particle vertex functions, we
pass on to the 1PI local functional by means of a Legendre transform. For the sake of conciseness, we have considered
in this work the two typical approximations for the diagrammatic methods: (i) the restriction to the local two-particle
vertices and (ii) the ladder approximation for the self-energy. With these assumptions, we could show how the ladder
self-energy diagrams generated by the 1PI approach also include contributions from local one-particle reducible
three-particle vertices, which, in the DF approach, can only be generated when explicitly computing the local
three-particle vertex. Hence, when adopting the usual approximations, the 1PI approach contains a {\sl larger set} of
diagrams than DF.

Let us also stress that the 1PI approach prevents the generation of spurious ``reducible diagrams''
present in the DF self-energy when restricting oneself to the two-particle vertices \cite{Katanin12}.
In this respect, the 1PI approach can be further used for a consistent formulation of the renormalization of the DF approach,
restricted to the two-particle level (e.g., within the functional renormalization-group analysis, Ref. \cite{fRGReview}).

As for the comparison with the D$\Gamma$A, we note that its derivation is purely based on diagrammatic considerations rather than on a path-integral formulation. However, when comparing the diagrams defining the ladder approximation of D$\Gamma$A with the corresponding ones from the 1PI approach, it turns out that they have a similar structure. We observe that - like for the DF approach - the 1PI method allows us to treat diagrams which are neglected in the ladder D$\Gamma$A
analysis, but are present in its parquet generalizations.

Beyond a detailed analysis of the diagrammatics of the 1PI, DF and D$\Gamma$A, we have also compared the numerical
results of the three approaches. For non-self-consistent calculations DF yields substantially weaker corrections to
DMFT than D$\Gamma$A and 1PI. Self-consistent results, which in the case of 1PI and D$\Gamma$A are mimicked by a
Moriyaesque $\lambda$ correction, are more similar. Here, we observe the general trend that 1PI yields somewhat
stronger corrections to the DMFT self-energy than D$\Gamma$A  at intermediate-to-strong coupling, yielding
results, which are close to those in DF approach. At weak-coupling we find the nonlocal corrections to the self-energy
in the 1PI approach to be smaller, than those in the D$\Gamma$A and DF approaches. We trace this back to the additional Feynman diagrams of 1PI which, in
comparison to D$\Gamma$A, substitute a bare interaction $U$ by a local vertex. At weak-coupling, this local vertex is
smaller than $U$ because of metallic screening processes. At strong-coupling it is larger, because of the formation of
a local moment associated to strong spin fluctuations at the MIT.

In summary, the 1PI approach unifies features of the DF and D$\Gamma$A approaches. Restricting ourselves to
(i) a truncation of the approaches at the two-particle local vertex level and (ii) ladder diagrams generated from these, the
1PI allows for a treatment of the nonlocal self-energy effects, accounting for the
non-ladder scattering processes.

\section{Acknowledgments}

We gratefully acknowledge financial support through the Austrian-Russian joint project from the Austrian Science Fund
(FWF), project I-610-N16 (GR, AT) and Russian Fund of Basic Research (RFBR), grant 10-02-91003-ANFa (AK), 13-02-00050 (VA);
the Program of the Russian Academy of Science Presidium Quantum microphysics of condensed matter 12-P-2-1017,
12-M-23-2020 and the grant of the Ministry of education and science of Russia No. 14.18.21.0076 (VA); as well as the
European Research Council under the European Union's Seventh Framework Programme (FP/2007-2013)/ERC through grant
agreement n. 306447 (KH). The numerical calculations presented in this paper have been performed on the Vienna
Scientific Cluster (VSC).

\section{Appendix A. 1PI transformations of the generating functional}

\subsection*{Calculation of \boldmath{$J[\phi^+,\phi]$}}
In order to pass to the 1PI functional, we change variables of integration $%
\widetilde{c}^{+},\widetilde{c}$ to the Legendre transformed quantities $%
\phi^{+},\phi$ [see Eq. (\ref{legendre}) and the definition of $\widetilde{\eta}$ below]:
\begin{equation}
 \label{legendre1}
 \begin{split}
 &\widetilde{c}_{k\sigma}^+=-\frac{\delta\Gamma_{\text{DMFT}}}{\delta\phi_{k\sigma}}-\eta_{k\sigma}^+ \\&
 \widetilde{c}_{k\sigma}=\frac{\delta\Gamma_{\text{DMFT}}}{\delta\phi_{k\sigma}^+}-\eta_{k\sigma}.
 \end{split}
\end{equation}
Considering that $\zeta_{\nu}$ and the source field $\eta^+$ and $\eta$ do not depend on $\widetilde{c}^+$ and
$\widetilde{c}$, the corresponding matrix $M[\phi^+,\phi]$ of this transformation can be written as:
\begin{equation}
 \label{equ:jacobimatrix}
  d\left(
\begin{array}{c}
\widetilde{c}_{k\sigma}^+\\
\widetilde{c}_{k\sigma}%
\end{array}%
\right)=
\underset{M[\phi^+,\phi]}{\underbrace{
\left(
\begin{array}{cc}
-\frac{\delta^2\Gamma_{\text{DMFT}}}{\delta\phi_{k'\sigma'}^+\delta\phi_{k\sigma}}&
-\frac{\delta^2\Gamma_{\text{DMFT}}}{\delta\phi_{k'\sigma'}\delta\phi_{k\sigma}}\\
\frac{\delta^2\Gamma_{\text{DMFT}}}{\delta\phi_{k'\sigma'}^+\delta\phi_{k\sigma}^+}&
\frac{\delta^2\Gamma_{\text{DMFT}}}{\delta\phi_{k'\sigma'}\delta\phi_{k\sigma}^+}
\end{array}%
\right)}} d\left(
\begin{array}{c}
\phi _{k'\sigma^{\prime }}^+\\
\phi _{k'\sigma ^{\prime }}%
\end{array}%
 \right).
\end{equation}
The calculation of the first (which will be needed later) and the second functional derivatives of the functional $\Gamma_{\text{DMFT}}[\phi^+,\phi]$ with respect to the fields $\phi^+$ and $\phi$ can be performed straightforwardly using the explicit expression for $\Gamma_{\text{DMFT}}$ given in Eq. (\ref{GammaDMFT}). The results are:
\begin{align}
 \label{gammafirstderivative}
 &\frac{\delta \Gamma _{\text{DMFT}}[\phi^+,\phi]}{\delta \phi_{k\sigma}}=\frac{1}{\beta}G_{\text{loc},\nu}^{-1}\phi^+_{k\sigma}-\frac{1}{\beta^3}\sum_{k_1q}\sum_{\sigma_1}\widetilde{\Gamma}_{\text{loc},\sigma\sigma_1}^{\nu_1\nu\omega}\phi^+_{k+q,\sigma}\phi^+_{k_1\sigma_1}\phi_{k_1+q,\sigma_1},\\
 &\frac{\delta \Gamma _{\text{DMFT}}[\phi^+,\phi]}{\delta \phi^{+}_{k\sigma}}=-\frac{1}{\beta}G_{\text{loc},\nu}^{-1}\phi_{k\sigma}+\frac{1}{\beta^3}\sum_{k_1q}\sum_{\sigma_1}\widetilde{\Gamma}_{\text{loc},\sigma\sigma_1}^{\nu\nu_1\omega}\phi_{k+q,\sigma}\phi^+_{k_1+q,\sigma_1}\phi_{k_1\sigma_1},
\end{align}
for the first derivatives and
\begin{equation}
 \label{equ:2ndderivatives}
 \begin{split}
 &\frac{\delta^2\Gamma_{\text{DMFT}}}{\delta\phi_{k'\sigma'}\delta\phi_{k\sigma}^+}=-\frac{1}{\beta}G_{\text{loc},\nu}^{-1}\delta_{kk'}\delta_{\sigma\sigma'}-\frac{1}{\beta^3}\sum_{q}\widetilde{\Gamma}_{\text{loc},\sigma\sigma'}^{\nu\nu'\omega}\phi_{k'+q,\sigma'}^+\phi_{k+q,\sigma}+\frac{\delta_{\sigma\sigma'}}{\beta^3}\sum_{q,\sigma_1}\widetilde{\Gamma}_{\text{loc},\sigma\sigma_1}^{\nu,\nu+\omega,\nu'-\nu}\phi_{k'+q,\sigma_1}^+\phi_{k+q,\sigma_1}\\
 &\frac{\delta^2\Gamma_{\text{DMFT}}}{\delta\phi_{k'\sigma'}^+\delta\phi_{k\sigma}^+}=-\frac{1}{\beta^3}\sum_{q}\widetilde{\Gamma}_{\text{loc},\sigma\sigma'}^{\nu,\nu'-\omega,\omega}\phi_{k+q,\sigma}\phi_{k'-q,\sigma'}\\
 &\frac{\delta^2\Gamma_{\text{DMFT}}}{\delta\phi_{k'\sigma'}\delta\phi_{k\sigma}}=\frac{1}{\beta^3}\sum_{q}\widetilde{\Gamma}_{\text{loc},\sigma\sigma'}^{\nu'-\omega,\nu\omega}\phi_{k'-q,\sigma'}^+\phi_{k+q,\sigma}^+\\
&\frac{\delta^2\Gamma_{\text{DMFT}}}{\delta\phi_{k'\sigma'}^+\delta\phi_{k\sigma}}=
\frac{1}{\beta}G_{\text{loc},\nu}^{-1}\delta_{kk'}\delta_{\sigma\sigma'}+\frac{1}{\beta^3}\sum_{q}\widetilde{\Gamma}_{\text{loc},\sigma\sigma'}^{\nu\nu'\omega}\phi_{k+q,\sigma}^+\phi_{k'+q,\sigma'}-\frac{\delta_{\sigma\sigma'}}{\beta^3}\sum_{q,\sigma_1}\widetilde{\Gamma}_{\text{loc},\sigma\sigma_1}^{\nu+\omega,\nu,\nu'-\nu}\phi_{k+q,\sigma_1}^+\phi_{k'+q,\sigma_1}.
 \end{split}
 \end{equation}
for the second functional derivatives. $\widetilde{\Gamma}_{\text{loc},\sigma\sigma'}^{\nu\nu'\omega}$ is defined below Eq. (\ref{GammaDMFT}).

Next, we single out the factor $\left[\beta G_{\text{loc},\nu}\right]^{-1}$ from the Jacobian
$M[\phi^+,\phi]=\left[\beta G_{\text{loc},\nu}\right]^{-1}\widetilde{M}[\phi^+,\phi]$ and omit it since it depends neither on the source fields $\eta^+$ and $\eta$ nor on the
integration variables $\phi^+$ and $\phi$ and, hence, does not contribute to
the derivatives of $\log Z$ w.r.t.\ the source fields (see the discussion in section III). Furthermore, we represent
$\widetilde{M}[\phi^+,\phi]=\mathds{1}+\widetilde{\cal M}[\phi^+,\phi]$ where
\begin{align}
 \label{equ:defD}
 &\widetilde{\cal M}[\phi^+,\phi]=-\frac{1}{\beta^2}G_{\text{loc},\nu}\times\nonumber\\&\times\sum_{q}
  \left(
   \begin{array}{c|c}
   \widetilde{\Gamma}_{\text{loc},\sigma\sigma'}^{\nu\nu'\omega}\phi_{k+q,\sigma}^+\phi_{k'+q,\sigma'}+&\multirow{2}{*}{$-\widetilde{\Gamma}_{\text{loc},\sigma\sigma'}^{\nu,\nu'-\omega,\omega}\phi_{k+q,\sigma}^+\phi_{k'-q,\sigma'}^+$}\\-\delta_{\sigma\sigma'}\sum_{\sigma_1}\widetilde{\Gamma}_{\text{loc},\sigma\sigma_1}^{\nu,\nu+\omega,\nu'-\nu}\phi_{k+q,\sigma_1}^+\phi_{k'+q,\sigma_1}&\\ \hline
   \multirow{2}{*}{$\widetilde{\Gamma}_{\text{loc},\sigma\sigma'}^{\nu,\nu'-\omega,\omega}\phi_{k+q,\sigma}\phi_{k'-q,\sigma'}$} &
   \widetilde{\Gamma}_{\text{loc},\sigma\sigma'}^{\nu\nu'\omega}\phi_{k'+q,\sigma'}^+\phi_{k+q,\sigma}
   +\\&-\delta_{\sigma\sigma'}\sum_{\sigma_1}\widetilde{\Gamma}_{\text{loc},\sigma\sigma_1}^{\nu,\nu+\omega,\nu'-\nu}\phi_{k'+q,\sigma_1}^+\phi_{k+q,\sigma_1}
  \end{array}\right),
\end{align}
The inverse (note that we are dealing with Grassmann integrals \cite{negeleorland}) of $J[\phi^+,\phi]$ is now given by
\begin{equation}
 \label{equ:defJ}
 J^{-1}[\phi^+,\phi]=\det \widetilde{M}[\phi^+,\phi]
\end{equation}
Here, $\det$ denotes the determinant w.r.t. the $k$ and $\sigma$ indices.
In order to include $J$ in the effective action, we transfer it to the exponent by taking its logarithm and make use of the general identity $\log\det A=\text{Tr}\log A$. Hence, we finally arrive at
\begin{equation}
 \label{equ:Jfinal}
 \log J[\phi^+,\phi]=-\text{Tr}\log\widetilde{M}[\phi^+,\phi]=-\text{Tr}\log\left(\mathds{1}+\widetilde{\cal M}[\phi^+,\phi]\right) .
\end{equation}
where Tr denotes the trace w.r.t. the $k$ and $\sigma$ indices.
Performing a Taylor expansion of the logarithm in the last term, we obtain an expansion of the Jacobian in $\phi^+$, $\phi$ fields.

In the {\sl first (quadratic) order} in fermionic fields we obtain the term with the structure $G_{\text{loc}}\Gamma_{\text{loc}}\phi^+\phi$. Hence, it corresponds to the first diagram in Fig. \ref{fig:diagramelements1pi}b and cancels the corresponding ones which are generated by the perturbation expansion of the 1PI functional Eq. (\ref{Eq1PI}). The terms of the {\sl second (quartic) order} can be schematically written as $\Gamma_{\text{loc}}G_{\text{loc}}^2\Gamma_{\text{loc}}(\phi^+\phi)(\phi^+\phi)$ and correspond to the second diagram in Fig. \ref{fig:diagramelements1pi}b.
Let us also note that higher order contributions in $\phi$, i.e., $O((\phi^{+}\phi)^3)$, generate terms that cancel the nonlocal corrections to the self-energy stemming from the three- (and more-)particle local 1PI vertices that are already taken into account at the two-particle vertex level via combination of the elements of diagram technique of
Fig. 3. In this way any possible double counting is avoided in the 1PI approach. For a more detailed discussion of this issue we refer to \cite{phdrorhinger}.

\subsection*{Transformation of integral variables and decoupling of the three-particle term}

In this section we decouple the term in the second line of Eq. (\ref{Eq1PI}), which contains a three-particle interaction, as discussed below Eq. (\ref{GammaDMFT}). For this purpose we consider the following Hubbard-Stratonovich transformations:
\begin{align}
 \label{hsdecoupling2}
 &\exp  \left\{\!
\beta\left( \frac{\delta \Gamma _{\text{DMFT}}[\phi^+,\phi]}{%
\delta \phi _{k\sigma }}\!+\!\eta _{k\sigma }^{+}\!\right) [\zeta_{\nu}^{-1}\!-\!G_{0k}^{-1}]^{-1}\left(\! -\frac{\delta \Gamma _{\text{DMFT}}[\phi^+,\phi]}{\delta \phi _{k\sigma }^{+}}\!+\!\eta _{k\sigma }\!\right) \right\}=\nonumber\\[0.2cm]=
 \int& d\psi_{k\sigma}^+d\psi_{k\sigma}
 \;\exp\biggl\{-\frac{1}{\beta}\left[\zeta^{-1}_{\nu}-G_{0k}^{-1}\right]\psi_{k\sigma}^+\psi_{k\sigma}\biggr\}\times\nonumber\\&\hspace{1.3cm}\times \exp\left\{\left[\left(\frac{\delta\Gamma_{\text{DMFT}}[\phi^+,\phi]}{\delta\phi_{k\sigma}}+\eta_{k\sigma}^+\right)\psi_{k\sigma}+\psi_{k\sigma}^+\left(-\frac{\delta\Gamma_{\text{DMFT}}[\phi^+,\phi]}{\delta\phi^+_{k\sigma}}+\eta_{k\sigma}\right)\right]\right\},
 \end{align}
where we neglected the prefactor $\beta\left[\zeta_{\nu}^{-1}-G_{0k}^{-1}\right]^{-1}$ in front of the functional integral in this equation, since it drops out in the calculation of the Green's function. In the next step we insert Eq. (\ref{hsdecoupling2}) into Eq. (\ref{Eq1PI}) and then perform the following shift of integration variables:
\begin{align}
 &\psi_{k\sigma}^+\rightarrow\psi_{k\sigma}^++\phi_{k\sigma}^+,&&\psi_{k\sigma}\rightarrow\psi_{k\sigma}+\phi_{k\sigma}.
\end{align}
One observes that the terms $\left(\delta\Gamma_{\text{DMFT}}/\delta\phi\right)\phi$ and $\phi^+\left(\delta\Gamma_{\text{DMFT}}/\delta\phi^+\right)$ in Eq. (\ref{Eq1PI}) are canceled by the corresponding ones from Eq. (\ref{hsdecoupling2}). Hence, one arrives at the following expression for the generating functional $Z[\eta^+,\eta]$:
\begin{align}
 \label{zdecouple}
 Z[\eta^+,\eta]=\int D[\phi^+,\phi]\;\exp\left.\biggl\{\right.&\left.-\frac{1}{\beta}\sum_{k,\sigma}\left[\zeta^{-1}_{\nu}-G_{0k}^{-1}\right]\left(\psi^+_{k\sigma}+\phi^+_{k\sigma}\right)\left(\psi_{k\sigma}+\phi_{k\sigma}\right)+\right.\nonumber\\&\left.+\frac{\delta\Gamma_{\text{DMFT}}[\phi^+,\phi]}{\delta\phi_{k\sigma}}\psi_{k\sigma}-\psi^+_{k\sigma}\frac{\delta\Gamma_{\text{DMFT}}[\phi^+,\phi]}{\delta\phi^+_{k\sigma}}-\Gamma_{\text{DMFT}}[\phi^+,\phi]+\right.\nonumber\\ &\left.+\eta^+_{k\sigma}\left(\psi_{k\sigma}-\phi_{k\sigma}\right)+\left(\psi^+_{k\sigma}-\phi^+_{k\sigma}\right)\eta_{k\sigma}\right.\biggr\}J[\phi^+,\phi].
\end{align}
Inserting now the explicit expressions for $\Gamma_{\text{DMFT}}[\phi^+,\phi]$ from Eq. (\ref{GammaDMFT}) and $(\delta\Gamma_{\text{DMFT}}/\delta\phi^{(+)}$ from Eq. (\ref{gammafirstderivative}) into Eq. (\ref{zdecouple}) one arrives at the final expression for the generating functional $Z[\eta^+,\eta]$ in the 1PI representation as given in Eq. (\ref{Zf}) [consider that $-\zeta^{-1}_{\nu}+G_{0k}^{-1}+G_{\text{loc},\nu}^{-1}=G_k^{-1}$].


\begin{thebibliography}{1}
\bibitem{noteJanis} We note, however, that in the diagrammatic approach to the Anderson localization of Refs. \cite{Janis1,Janis2}, by taking the  infinite dimensional limit corresponding to  DMFT the exact solution reduces to the Coherent Potential Approximation (CPA). This simplification, of course, does not apply to the case of the Hubbard model problems considered in the present work.
\bibitem{dgadfdifference}
Replacing $G_0$ by the (interacting) DMFT Green's function $G$ and substituting
correspondingly  the ``Weiss''-field (i.e., the non-interacting local Green's function) $\zeta$ by its interacting counterpart $G_{\text{loc}}$ leads to exactly the same decoupling scheme as used in DF\cite{DF}. This can be easily seen from the identity
$\zeta^{-1}_{\nu}-G_{0k}^{-1}=G_{\text{loc},\nu}^{-1}-G_k^{-1}$.
However, we think that the way of splitting local and nonlocal degrees of freedom as performed in Eq. (\ref{Split}) is physically more transparent.
\bibitem{note_scDF}  We recall that in DF self-consistency can be performed at two-different levels: (i) by re-inserting the Green's functions of the dual electrons in the self-energy diagrams of the dual electrons (``inner self-consistency'') and (ii) by readapting the AIM to take into account that the local physics has also changed due to the nonlocal DF corrections (``outer self-consistency'').
\bibitem{note_scDGA} In D$\Gamma$A, as well in 1PI, one can correct for the non-self consistency effects appearing for the two-particle quantities in one-shot calculations for the ladder-diagrams, by correcting the two-particle susceptibilities within the Moriya schemes, as explained in Ref. \cite{DGA2}.
\bibitem{1pi_selfconsistency} For the 1PI approach the self-consistency could be alternatively  achieved as in DF, i.e., by redefining the local problem $\zeta(i\omega_n)$, so to minimize the corrections w.r.t. the new local AIM.
\end{thebibliography}
\end{document}